\pdfoutput=1
\def\paperchapt/{paper}

\documentclass[journal,twoside]{IEEEtran}

\usepackage{flushend}  
\usepackage{ifpdf}
\usepackage{cite}
\ifCLASSINFOpdf
  \usepackage[pdftex]{graphicx}
  \DeclareGraphicsExtensions{.pdf,.jpeg,.png}
\else
  \usepackage[dvips]{graphicx}
  \DeclareGraphicsExtensions{.eps}
\fi
\graphicspath{{./figures/}}
\usepackage[cmex10]{amsmath}
\usepackage{amsfonts}  
\usepackage{amssymb}  
\usepackage{amsthm}
\usepackage{array}
\usepackage{dcolumn}
\usepackage[caption=false,font=footnotesize]{subfig}
\usepackage{url}

\newcommand{\vect}[1]{\mathbf{#1}}
\newcommand{\matr}[1]{\mathbf{#1}}
\newcommand{\vsym}[1]{\boldsymbol{#1}}    

\newcommand{\set}[1]{\mathcal{#1}}

\newcommand{\F}{\mathbb{F}}
\newcommand{\R}{\mathbb{R}}
\newcommand{\C}{\mathbb{C}}

\DeclareMathOperator{\order}{order}
\DeclareMathOperator{\size}{size}

\DeclareMathOperator{\diag}{diag}

\DeclareMathOperator{\var}{VAR}
\DeclareMathOperator{\QF}{Q}
\newcommand{\vlambda}{\vsym{\lambda}}
\newcommand{\vLambda}{\vsym{\Lambda}}
\newcommand{\Pb}{P_\mathrm{b}}
\newcommand{\Pf}{P_\mathrm{f}}
\newcommand{\Eb}{E_\mathrm{b}}

\newcommand{\dv}{d_\mathrm{v}}
\newcommand{\dc}{d_\mathrm{c}}

\theoremstyle{plain}
\newtheorem{thm}{Theorem} 
\newtheorem{lem}[thm]{Lemma}

\theoremstyle{definition}
\newtheorem{defn}{Definition} 
\newtheorem{exmp}{Example} 
\newtheorem{ass}{Assumption} 

\theoremstyle{remark}

\makeatletter
\newlength \figwidth
\if@twocolumn
\else
\fi
\makeatother

\if@twocolumn  
  \renewcommand{\arraystretch}{0.92} 
\else
  \renewcommand{\arraystretch}{0.9} 
\fi

\newcommand{\mysmallarraydecl}{\renewcommand{%
\IEEEeqnarraymathstyle}{\scriptstyle}%
\renewcommand{\IEEEeqnarraytextstyle}{\scriptsize}%
\renewcommand{\baselinestretch}{0.93}%
\settowidth{\normalbaselineskip}{\scriptsize
\hspace{\baselinestretch\baselineskip}}%
\setlength{\baselineskip}{\normalbaselineskip}%
\setlength{\jot}{0.25\normalbaselineskip}%
\setlength{\arraycolsep}{3pt}}

\usepackage{hyperref}
\hypersetup{
  pdfinfo={
    Title={Error Floor Approximation for LDPC Codes in the AWGN Channel},
    Author={Brian K. Butler},
    Subject={Information Theory, Error Correction Coding},
    Keywords={LDPC codes; Error Correction; Error Floor; Message Passing Decoding; Near-codeword; Trapping Set; Absorbing Set; Linear Analysis}
  }
}

\begin{document}

\title{Error Floor Approximation for {LDPC} Codes in the AWGN Channel}

\author{Brian~K.~Butler,~\IEEEmembership{Senior~Member,~IEEE,}
        Paul~H.~Siegel,~\IEEEmembership{Fellow,~IEEE}
\thanks{This work was presented in part at the Forty-Ninth Annual Allerton Conference, Allerton House, Illinois, Sept 2011.}
\thanks{The authors are with the Department of Electrical and Computer Engineering,
University of California, San Diego (UCSD), La Jolla, CA 92093 USA (e-mail: butler@ieee.org, psiegel@ucsd.edu).}
\thanks{This work was supported in part by the Center for Magnetic Recording Research at UCSD and by the National Science Foundation (NSF) under Grant CCF-0829865 and Grant CCF-1116739.}
}

\maketitle

\ifCLASSOPTIONpeerreview
	\markboth{Error Floor Approximation for {LDPC} Codes in the AWGN Channel}%
	{Error Floor Approximation for {LDPC} Codes in the AWGN Channel}
\else
	\markboth{\MakeLowercase{submitted to} \textit{IEEE Transactions on Information Theory} \;\;\;\;Version:  June 4, 2013}%
	{Butler and Siegel: Error Floor Approximation for {LDPC} Codes in the AWGN Channel}
\fi
%

\begin{abstract} 

This \paperchapt/ addresses the prediction of error floors of low-density parity-check (LDPC) codes with variable nodes of constant degree in the additive white Gaussian noise (AWGN) channel.
Specifically, we focus on the performance of the sum-product algorithm (SPA) decoder formulated in the log-likelihood ratio (LLR) domain.
We hypothesize that several published error floor levels are due to the manner in which decoder implementations handled the LLRs at high SNRs.
We employ an LLR-domain SPA decoder that does not saturate near-certain messages and find the error rates of our decoder to be lower by at least several orders of magnitude.
We study the behavior of trapping sets (or near-codewords) that are the dominant cause of the reported error floors.

We develop a refined linear model, based on the work of Sun and others, that accurately predicts error floors caused by elementary tapping sets for saturating decoders.
Performance results of several codes at several levels of decoder saturation are presented.
\end{abstract}
%


%
\IEEEpeerreviewmaketitle

\section{Introduction} 
\IEEEPARstart{A}{} very important class of modern codes, the low-density parity-check (LDPC) codes,
was first published by Gallager in 1962 \cite{Gal62,Gal63}.
LDPC codes are linear block codes described by a sparse parity-check matrix.
Decoding algorithms for LDPC codes are generally iterative.
The renaissance of interest in these codes began with work
by MacKay, Neal, and Wiberg in the late 1990s \cite{MK96,Wiberg,WibergPaper}.
Progress has been rapid, with information-theoretic channel capacity essentially reached in some applications of LDPC codes
and standardization complete for their commercial use in others, \textit{e.g.}, DVB-S2 for satellite broadcast
and IEEE 802.3an for $10$ Gbit/s Ethernet.

The graph of an iteratively decoded code's error rate performance versus the channel quality is typically divided into two regions.
The first region, termed the \emph{waterfall} region, occurs at poorer channel quality, close to the decoding threshold, and is characterized by a rapid drop in error rate as channel quality improves.
The second region, called the \emph{error floor} region, is the focus of the present \paperchapt/.
The error floor appears at higher channel quality and is characterized by a more gradual decrease in error rate as channel quality improves.
For message-passing iterative decoders the error floor is apparently determined by small structures within the code that are specific to the selected graphical description of the code.

The understanding of the LDPC code error floor has progressed significantly, but issues remain.
For the binary erasure channel (BEC), the structures that limit the performance of message-passing iterative decoders as channel conditions improve are known as \emph{stopping sets} \cite{Di02}.
These sets have a combinatorial description and can be enumerated to accurately predict the error floor for the BEC.

Other memoryless channels are more difficult to characterize and the presence of error floors has limited the adoption of LDPC codes in some applications.
For example, in the DVB-S2 standard, an outer algebraic code is concatenated with an LDPC code to ensure a low error floor.
For Margulis-type codes, MacKay and Postol found that the error floors seen in the additive white Gaussian noise (AWGN) channel with sum-product algorithm (SPA) decoding were caused by \emph{near-codewords} in the Tanner graph \cite{MKfloor}.
Richardson wrote a seminal paper on the error floors of memoryless channels shortly afterward \cite{RichFloors}.
In it, he called near-codewords \emph{trapping sets} and defined them with respect to the decoder in use as the error-causing structures.
The parameters $(a, b)$ are used to characterize both near-codewords and trapping sets, where
$a$ is the number of variable nodes in the set and $b$ is the number of unsatisfied check nodes when just those variable nodes are in error.
The $(a, b)$ parameters of error-floor-causing structures are typically small.

In \cite{RichFloors}, Richardson emphasized error-floor analysis techniques for the AWGN channel .
He detailed a methodology to estimate a trapping set's impact on the error rate by simulating the AWGN channel in the neighborhood of the set.
His semi-analytical technique was shown to be accurate at predicting the error floors given that the trapping sets were known.
The method involved significant simulation, but much less than standard Monte Carlo simulation.
Roughly speaking, error floors can be measured down to frame error rates of
about $10^{-8}$ in standard computer simulation and about $10^{-10}$ in hardware simulation,
depending on code complexity and the computational resources available.
Richardson's method allows us to reach orders of magnitude further in characterizing the error floor.

Further work on trapping sets includes algorithms to enumerate the occurrence of small trapping sets 
in specific codes \cite{Poor,SCH10,Abu-Surra,XiaojieGC,Karimi} and the characterization of trapping set parameters in random ensembles of LDPC codes \cite{Milenk07,AbuDiv11}.

Several other significant works on error floors in the AWGN channel exist.
For example, Dolecek \textit{et al.} noted empirically that the trapping sets dominating the error floor, termed \emph{absorbing sets}, had a combinatorial description \cite{DolecekTIT}. 
Their study made use of hardware-oriented SPA decoders.
A variety of works describe techniques to overcome error floors.
With respect to these, we note that \cite{Han09} proposes several different techniques, 
\cite{Land} slows decoder convergence using averaged decoding, 
\cite{ZhangJSSC} selectively biases the messages from check nodes, 
\cite{VilaCasadoTC} employs informed scheduling of nodes for updating, 
\cite{Sharon} adds check equations to the parity-check matrix, and
\cite{DivCon} replaces traditional BP with a ``difference-map'' message passing algorithm.
In contrast to these, we present in this \paperchapt/ an error floor reduction technique that makes no changes to the standard SPA decoding aside from fixing common numerical problems.

Following the pioneering work of Richardson, several authors have proposed analytical techniques to predict error floors in AWGN. 
In his Ph.D. thesis \cite{SunPhD}, Sun developed a model of elementary trapping set behavior based on a linear state-space model of decoder dynamics on the trapping set, using density evolution (DE) to approximate decoder behavior outside of the trapping set \cite{SunPhD,SunAller}. \nocite{SunAller}
This work has gained little attention, even though it reaches a conclusion contrary to prior results.
Specifically, Sun's model shows no error floor for elementary trapping sets (excluding codewords) in regular, infinite-length LDPC
codes if the variable-degree of the code is at least three and the decoder's metrics
are allowed to grow very large. Sun is able to make similar claims
for irregular LDPC codes under certain conditions.
In these cases, Sun shows that the graph outside of the trapping set
will eventually correct the trapping set errors; in his example, using a $(3,1)$ trapping set,
this occurred at a mean log-likelihood ratio (LLR) of about $10^{10}$ after about $40$ iterations.

Sun's analytical development was asymptotic in the number of iterations and the block length of the code.
When applied to finite-length codes with cycles, we show that the model breaks down due to strongly correlated LLR messages when the decoder is non-saturating.

Schlegel and Zhang \cite{SCH10} extended Sun's work by adding time-variant gains
and channel errors from outside of the trapping set to the model.
For the 802.an code, using a model specific to the dominant $(8,8)$ trapping set, they then compared their analytically predicted error floors to error floors found by  FPGA-based simulation and importance sampling simulation. 
Their results show that the external errors have very little impact and that the error floor can be predicted quite accurately.  

We generalize their model and compare predicted error floors to simulations for four different codes with several LLR limits.
We find the model estimates the error floor to within $0.15$ dB for three of the four codes that we examine.
We also show how a model that breaks down for finite-length codes in a non-saturating scenario may work fairly well when saturated LLRs are introduced.

Brevik and O'Sullivan developed a new mathematical description of decoder convergence
on elementary trapping sets using a model of the probability-domain SPA decoder \cite{Osullivan}.
While considering just the channel inputs, they are able to find regions of convergence to the correct codeword,
regions of convergence to the trapping set, and
regions of non-convergence typified by an oscillatory behavior.

Xiao \textit{et al.} \cite{XiaoErrorEstim} have developed a technique to estimate the error floor in the quantized decoder case using a binary input AWGN channel.
In contrast, the present \paperchapt/ focuses on unquantized decoders and different floor-estimation techniques.

Several authors have found empirically that increasing LLR limits in the SPA decoder lowers the error floor \cite{ThorpeLowC,ChenFoss, Zhao, ZhangQuantFloor}.
Our findings concur with theirs. 
Furthermore, our results show that when care is taken to implement the SPA decoder without saturation, error floors can be lowered quite dramatically, often by as much as several orders of magnitude. 
In the cases we examine, the error floors are reduced by many orders of magnitude.
Recent independent work by Zhang and Schlegel confirms this \cite{SCH11,SCH12}.

\begin{figure}
\centering
\includegraphics[width=\figwidth]{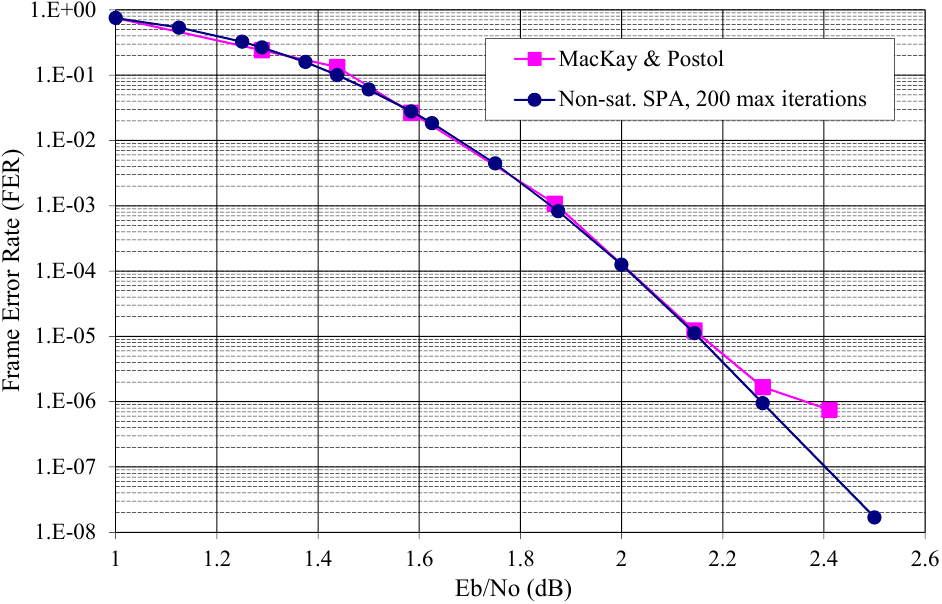}
\caption{FER vs. $\Eb/N_0$ for (2640,1320) Margulis LDPC code in AWGN.}
\label{fig_MFER}
\end{figure}

Fig.~\ref{fig_MFER} illustrates the improvement in performance that non-saturating decoders can provide.  
The curves represent the frame error rate (FER) reported in \cite{MKfloor} for the $(2640, 1320)$ Margulis code and the corresponding 
results obtained from our simulation of a non-saturating SPA decoder.
The error floor found in \cite{MKfloor} starts at an FER of about $10^{-6}$ with an $\Eb/N_0$ of $2.4$ dB. 
The dominant errors corresponded to near-codewords. 
(It should be noted that others have reported even higher error floors for this code, with an FER of $10^{-6}$ appearing at an $\Eb/N_0$ of $2.8$ dB \cite{RichFloors,Han09}.) 
Our simulation, on the other hand, shows no evidence of an error floor. 
The lowest simulated point at an FER of $1.8 \cdot 10^{-8}$ represents $154$ error events observed in $8600$ hours of floating-point simulation, with a maximum of $200$ iterations per decoded  frame. 
Just one of the error events was a $(14,4)$ near-codeword, which is one of the two dominant errors floor causing structures in the previously reported results. 
Thus, we see very little of the usual early signs that an error floor is developing, suggesting that any error floor would manifest itself only at a substantially lower FER. 

The framework we develop in this \paperchapt/ will help to explain the empirically observed error floor properties described above. 
Our focus will be on the prediction of error floors for binary LDPC codes with constant variable node degree when used on the AWGN channel.
The results presented here are an extension of those discussed in \cite{SunPhD,SCH10,SCH11,RichFloors}.
Section~\ref{sect-pre} introduces the general terminology related to nonnegative matrices, graphs, trapping sets, and SPA decoders.
We introduce message saturation as a technique to avoid numerical overflow in the SPA decoder and then describe the non-saturating decoding algorithm that will provide our performance benchmarks.
Section~\ref{sect-ss} develops the state-space model of decoder dynamics on the trapping set we will be using for the analytical results of this \paperchapt/.
It also develops approximations to the dominant eigenvalue required to characterize the system, using connections to graph theory.
Section~\ref{sect-domprob} develops the probability that the state-space model fails to converge to the correct decoding solution and the related error-rate estimates for the error floor.
Section~\ref{sect-analnosat} discusses the use of the Gaussian approximation for DE to model the unsatisfied-check LLR values to determine the ultimate error floor bounds of non-saturating decoders.
Section~\ref{sect-margnosat} applies a modified version of Richardson's semi-analytical technique to estimate the error floor of a non-saturating SPA decoder.
Section~\ref{sect-numer} demonstrates the accuracy of the state-space model's predictions of saturated decoder performance by comparing them to simulation results for four LDPC codes.
Finally, we draw our conclusions.
Several additional related topics are covered in the appendices.

\section{Preliminaries} 
\label{sect-pre}
\subsection{Nonnegative Matrices}
\label{ss-nm}
In this subsection we introduce the terminology of Perron-Frobenius theory of nonnegative matrices \cite{Horn,Meyer,Seneta,Minc}.

A vector $\vect{v} \in \R^{n}$ is said to be \emph{positive} if every element is strictly greater than zero.
Likewise, the vector $\vect{v}$ is said to be \emph{nonnegative} if every element is greater than or equal to zero.
In the present \paperchapt/, we use column vectors by default.

A matrix $\matr{M}\in \R^{m\times n}$ is said to be a \emph{positive matrix} if every entry is strictly greater than zero.
Likewise, the matrix $\matr{M}$ is said to be a \emph{nonnegative matrix} if every entry is greater than or equal to zero.
Moreover, the nonnegative matrix $\matr{M}$ is said to be a {$(0,1)$-matrix} if every entry belongs to the set $\{0,1\}$.

A \emph{permutation matrix} $\matr{P}$ is a square $(0,1)$-matrix, in which each row and each column contains a single $1$ entry.
The \emph{symmetric permutation} of the square matrix $\matr{M}$ by $\matr{P}$ is $\matr{P} \matr{M} \matr{P}^T$.
The nonnegative square matrix $\matr{M}$ is said to be a \emph{reducible matrix} if there exists a permutation matrix $\matr{P}$ such that
$\matr{P} \matr{M} \matr{P}^T= \bigl[\begin{smallmatrix}\matr{X}&\matr{Y}\\ \matr{0}&\matr{Z}\end{smallmatrix}\bigr]$
where $\matr{X}$ and $\matr{Z}$ are square submatrices.  
Otherwise, the matrix $\matr{M}$ is said to be an \emph{irreducible matrix}.

Given $\matr{M} \in \R^{m\times m}$, the set of distinct eigenvalues is called the \emph{spectrum} of $\matr{M}$, denoted
$\sigma(\matr{M}) = \left\{ \mu_1,\ldots,\mu_n \right\}$, where $\mu_k \in \C$.
The \emph{spectral radius} of $\matr{M}$ is the real number $\rho(\matr{M}) = \max \left\{ |\mu| : \mu \in \sigma(\matr{M}) \right\}$.
The \emph{spectral circle} is a circle in the complex plane about the origin with radius $\rho(\matr{M})$.

The following are a few important results from the theory of nonnegative matrices. 
Let $\matr{M}$ be an irreducible nonnegative matrix.
The matrix $\matr{M}$ has a simple, real eigenvalue $r$ on the spectral circle.
Also, there exists a positive vector $\vect{w}_1$ such that $\vect{w}_1^T\matr{M}=r \vect{w}_1^T$, which is the unique positive left eigenvector of $\matr{M}$ up to a positive scale factor.
If $\matr{M}$ has only one eigenvalue on the spectral circle, then it is said to be \emph{primitive}.
If $\matr{M}$ has $h>1$ eigenvalues with modulus equal to $\rho(\matr{M})$, then it is said to be \emph{imprimitive} or a cyclic matrix.
The value $h$ is known as the \emph{index of imprimitivity} or period.


\subsection{General Graph Theory}
\label{ss-ggt}
An \emph{undirected graph} $G=(V,E)$ consists of a finite set of vertices $V$ and a finite collection of edges $E$.
Each edge is an unordered pair of vertices $\{v_i,v_j\}$ such that the edge joins vertices $v_i$ and $v_j$.
Given the edge $\{v_i,v_j\}$, we say that vertex $v_i$ is \emph{adjacent} to vertex $v_j$, and \textit{vice versa}.
A vertex and edge are \emph{incident} with one another if the vertex is contained in the edge.

We do not give further consideration to graphs with self-loops.
A \emph{self-loop} is an edge joining a vertex to itself. 
Loopless graphs are commonly known as \emph{multigraphs}, which may contain parallel edges.
\emph{Parallel edges} are multiple inclusions of an edge in the edge collection.
A \emph{simple graph} has neither self-loops nor parallel edges.

The \emph{order} of a graph is the number of vertices and the \emph{size} is the number of edges.
The degree $d(v_i)$ of vertex $v_i$ is the number of edges incident with $v_i$.
Euler's handshaking lemma states that the sum of all degrees of a graph must be twice its number of edges.
It follows from the fact that each edge in the graph must contribute $2$ to the sum of all degrees.
Thus, in equation form for the undirected graph $G=(V,E)$, it may be stated that
\begin{alignat*}{2}
\order{(G)} =& \left| V \right|\quad\mbox{and}\\
\size{(G)} =& \left| E \right| = \frac{1}{2} \sum_{v_i \in V} d(v_i).
\end{alignat*}
A \emph{regular} graph is a graph whose vertices are all of equal degree.

In an undirected graph $G$, a \emph{walk} between two vertices is an alternating sequence of incident vertices and edges.
The vertices and edges in a walk need not be distinct.
The number of edges in a walk is its \emph{length}.
The vertices $v_i$ and $v_j$ are said to be \emph{connected} if the graph contains a walk of any length from $v_i$ to $v_j$, 
noting that every vertex is considered connected to itself.
A graph is said to be \emph{connected} if every pair of vertices is connected, otherwise the graph is said to be \emph{disconnected}.
The vertices of a disconnected graph may be partitioned into connected components.

{Unique edges} are those edges which appear only once in the edge collection of the graph. 
A walk that \emph{backtracks} is a walk in which a unique edge appears twice or more in-a-row in the walk.

A \emph{closed walk} is a walk that begins and ends on the same vertex.
A \emph{cycle} is a closed walk with no repeated edges or vertices (except the initial and final vertex) of length at least two.
The \emph{girth} of a graph is the length of its shortest cycle, if it has cycles.
If every vertex in a multigraph has at least degree two, then the graph contains one or more cycles.

A \emph{leaf} is a vertex of degree one.
A \emph{tree} is a connected graph without cycles.
Trees, with at least two vertices, have leaves.
We call an undirected graph \emph{leafless} if it does not contain leaves.

We will say that two graphs are \emph{identical} if they have equal vertex sets and equal edge collections.
Two graphs are said to be \emph{isomorphic} if there exists between their vertex sets a
one-to-one correspondence having the property that whenever two vertices are adjacent
in one graph, the corresponding two vertices are adjacent in the other graph.
This correspondence relationship is called an \emph{isomorphism}.
The isomorphism is a relabeling of the vertices that preserves the graph's structure.

The \emph{adjacency matrix} $\matr{A}(G)$ of any simple graph $G$ of order $n$ is the $n\times n$ symmetric $(0,1)$-matrix
whose $(i,j)$ entry is $1$ if and only if $\{v_i,v_j\}$ is an edge of $G$.
More generally, the adjacency matrix $\matr{A}(G')$ of any multigraph $G'$ is the symmetric nonnegative matrix
whose $(i,j)$ entry indicates the number of edges joining $v_i$ and $v_j$ in $G'$.
The number of walks of exactly length $p$ between vertices $v_i$ and $v_j$, in graph $G$, is the $(i,j)$ entry of $\matr{A}(G)^p$.
Two graphs 
are isomorphic if and only if there exists a symmetric permutation that relates their adjacency matrices.

The \emph{incidence matrix} $\matr{N}(G)$ of any multigraph $G$ of order $n$ and size $m$ is the $n\times m$ $(0,1)$-matrix
whose $(i,j)$ entry is $1$ if and only if the $i$th vertex is incident with $j$th edge of $G$.
It then follows that the multigraph's adjacency matrix may be stated as
$$
\matr{A}(G)=\matr{N}(G) \matr{N}(G)^T - \diag{(d(v_1), d(v_2), \ldots , d(v_n))}.
$$

A \emph{cycle graph}, denoted $C_n$, is a graph of order $n \ge 2$, in which the distinct vertices of the set $\{v_1,v_2,\ldots,v_n\}$
are joined by the edges from the collection $\{\{v_1,v_2\},$ $\ldots, \{v_{n-1},v_n\},$ $\{v_n,v_1\}\}$.
A cycle graph is a single cycle.

A \emph{bipartite graph} $B=(V,C,E)$ is a special case of an undirected graph in which 
the graph's vertices may be partitioned into two disjoint sets $V$ and $C$.
Each edge $e\in E$ of a bipartite graph joins a vertex from $V$ to a vertex from $C$. 
Hence, bipartite graphs cannot have self-loops by definition.
The parity-check matrix $\matr{H}$ over $\F_2$ describes a bipartite graph $B$, called the \emph{Tanner graph} of $\matr{H}$,
in which the vertices are known as variable nodes $V$ and check nodes $C$.
Tanner graphs of binary codes do not have parallel edges.
The parity-check matrix $\matr{H}$ is related to the adjacency matrix $\matr{A}(B)$ of the Tanner graph $B$ as
\begin{equation}
\label{bipart}
\matr{A}(B) = \begin{bmatrix}
\matr{0}&\matr{H}^T\\
\matr{H}&\matr{0}%
\end{bmatrix}.
\end{equation}

A $\dv$\emph{-variable-regular} Tanner graph is a Tanner graph whose variable nodes all have equal degree $\dv$, and
a $(\dv,\dc)$ \emph{regular} Tanner graph is both $\dv$-variable-regular and has check nodes all of degree $\dc$.  

Given a subset of the variable nodes $\set{S} \subseteq V$, we use $\set{N}(\set{S})$ to denote the set of adjacent check nodes.
Given a Tanner graph $B=(V,C,E)$ and any set of variable nodes $\set{S} \subseteq V$, let $B_\set{S}$ represents the \emph{induced subgraph} of $\set{S}$.
That is, $B_\set{S}=(\set{S},\set{N}(\set{S}),E_\set{S})$ is a bipartite graph containing the variable
nodes $\set{S}$, the check nodes $\set{N}(\set{S})$, and the edges between them $E_\set{S}$.
We will frequently refer to these induced subgraphs as Tanner subgraphs with parity-check submatrix $\matr{H}_\set{S}$.
The submatrix $\matr{H}_\set{S}$ is formed by selecting the columns of $\matr{H}$ as indexed by the members of the set $\set{S}$
and optionally removing any resulting all-zero rows.

The \emph{line graph} $\mathcal{L}(G)$ of a multigraph $G$ is the graph whose vertices are the edges of $G$. 
Two vertices of $\mathcal{L}(G)$ are adjacent if and only if their corresponding edges in $G$ have a vertex (or two) in common.
In fact, two parallel edges connect vertices in $\mathcal{L}(G)$ if and only if their corresponding edges in $G$ are parallel.
We find that the line graph's adjacency matrix is related to the incidence matrix of $G$ by $\matr{A}(\mathcal{L}(G))=\matr{N}(G)^T \matr{N}(G) - 2\,\matr{I}$.

For the line graph $\mathcal{L}(G)$ of multigraph $G=(V,E)$, 
\begin{alignat}{2}
\label{orderLG}
\order{(\mathcal{L}(G))} =& \left| E \right|=\frac{1}{2} \sum_{v_i \in V} d(v_i)\quad\mbox{and}\\
\label{sizeLG}
\size{(\mathcal{L}(G))} =& \frac{1}{2} \sum_{v_i \in V} d(v_i) \left(d(v_i)-1 \right) \\
\notag
=& \frac{1}{2} \sum_{v_i \in V} d(v_i)^2 -|E|.
\end{alignat}

The \emph{spectral radius} $\rho(G)$ of a multigraph $G=(V,E)$ is defined to be the spectral radius of the matrix $\matr{A}(G)$, and it is bounded by
\begin{equation}
\label{genspecG}
\frac{1}{|V|} \sum_{v_i \in V} d(v_i) 
\le  \rho(G)  \le
\max_{v_i \in V} d(v_i).
\end{equation}
Note that $d(v_i)$ is just the $i$th row (or column) sum of $\matr{A}(G)$. 
Since the matrix $\matr{A}(G)$ is symmetric, the lower bound follows by applying the Rayleigh quotient to $\matr{A}(G)$ with an all-one vector \cite[pp.~176--181]{Horn}.
The upper bound applies since the matrix $\matr{A}(G)$ is nonnegative, and is due to Frobenius \cite[p.~492]{Horn}.
The lower bound holds with equality if and only if the all-one vector is an eigenvector of $\matr{A}(G)$ corresponding to the eigenvalue $\rho(G)$.
If $G$ is connected, then the upper bound holds with equality if and only if $G$ is regular.
When $G$ is connected these conditions are equivalent.

A \emph{directed graph} or \emph{digraph} $D=(Z,A)$ consists of a set of vertices $Z$ and a collection of directed edges or \emph{arcs} $A$.
Each arc is an ordered pair of vertices $(z_i,z_j)$ such that the arc is directed from vertex $z_i$ to vertex $z_j$.
For arc $(z_i,z_j)$, we call the first vertex $z_i$ its \emph{initial vertex} and the second vertex $z_j$ its \emph{terminal vertex}.
We will focus on \emph{simple digraphs} which exclude self-loops and parallel arcs. 

In a digraph $D$, a \emph{directed walk} is an alternating sequence of vertices and arcs from $z_i$ to $z_j$ in $D$ such that every
arc $a_i$ in the sequence is preceded by its initial vertex and is followed by its terminal vertex.
A digraph $D$ is said to be \emph{strongly connected} if for any ordered pair of distinct vertices $(z_i,z_j)$
there is a directed walk in $D$ from $z_i$ to $z_j$. 
For example, the digraph in Fig.~\ref{fig42digraph} is strongly connected.

The \emph{adjacency matrix} $\matr{A}(D)$ of any simple digraph $D$ is the $(0,1)$-matrix whose $(i,j)$ entry is $1$ if and only if $(z_i,z_j)$ is an arc of $D$.

\begin{lem}
\label{Lconn}
A $(0,1)$-matrix is irreducible if and only if the associated digraph is strongly connected.
\end{lem}
\begin{IEEEproof}
See \cite[p.~78]{Minc}.
\end{IEEEproof}

The \emph{outdegree} $d^+_i$ of vertex $z_i$ is the number of arcs in digraph $D$ with initial vertex $z_i$.
Likewise, the \emph{indegree} $d^-_i$ of vertex $z_i$ is the number of arcs with terminal vertex $z_i$.
For the digraph $D=(Z,A)$, we note that
\begin{alignat*}{2}
\order{(D)} =& \left| Z \right|\quad\quad\mbox{and}\\
\size{(D)} =& \left| A \right| = \sum_{z_i \in Z} d^+_i = \sum_{z_i \in Z} d^-_i.
\end{alignat*}
A \emph{regular} digraph is a digraph whose vertices are all of equal indegree and outdegree.
The \emph{spectral radius} $\rho(D)$ of a digraph $D$ is defined to be the spectral radius of its matrix $\matr{A}(D)$, and it is bounded in the following lemma.

\begin{lem}
\label{lemFrob1}
Let the $i$th vertex $z_i$ of digraph $D=(Z,A)$ have outdegree $d^+_i$.
Then the spectral radius $\rho(D)$ of $D$ is bounded above and below by
\begin{equation}
\label{genspecD}
\min_{z_i \in Z} d^+_i
\le  \rho(D)  \le
\max_{z_i \in Z} d^+_i.
\end{equation}
If $D$ is strongly connected, then the inequalities are strict unless the digraph has regular outdegree.
Analogous statements hold for the indegrees.
\end{lem}
\begin{IEEEproof}
See \cite[p.~492]{Horn} and \cite[pp.~8 and 22]{Seneta}.
\end{IEEEproof}


The interested reader is referred to \cite{Minc,Trudeau,BrualdiRyser,CveRow}, etc., for a more complete treatment of the subject.
Our use of graph theory has parallels to \cite{Osullivan}. 

\subsection{LDPC Codes, the AWGN Channel, and SPA Decoding}
\label{ss-awgnspa}
LDPC codes are defined by the null space of a parity-check matrix $\matr{H}$.
The codewords are the set of column vectors $\set{C}$, such that $\vect{c} \in \set{C}$ satisfies 
$\matr{H} \vect{c} = \vect{0}$ over a particular field.
A given code can be described by many different parity-check matrices.

The $\matr{H}$ matrix over $\F_2$ may be associated with a bipartite graph $B=(V,C,E)$ termed a {Tanner graph} described in the previous subsection.
The set of variable nodes $V$ represent the symbols of the codeword that are sent over the channel and correspond to the columns of the parity-check matrix.
The set of check nodes $C$ enforce the parity-check equations represented by the rows of $\matr{H}$.

\begin{ass}
We are only concerned with LDPC codes over the binary field $\F_2$
and with binary antipodal signaling over the AWGN channel.
\end{ass}

The SPA decoder would be optimal with respect to minimizing the symbol error rate, if the Tanner graph had no cycles.
In any ``good" finite-length code, the Tanner graph will indeed have cycles \cite{cyclefree}, but we generally find that the SPA decoder does quite well.
Like many, we prefer to implement our decoder simulation in the LLR-domain, as it is close to the approximations typically used in building hardware.
Since we will assume that the reader is familiar with SPA decoding of LDPC codes over the AWGN channel (see \cite[\textsection~5.4]{RyanLin} and \cite{Kschischang}), the remainder of this subsection is presented merely to clarify notation.

We take the channel SNR to be $1/\sigma^2 = 2 R \Eb / N_0$.
We denote the intrinsic channel LLR for the $i$th received symbol by $\lambda_i$, and perform check node updates according to
\begin{equation}
\label{CNupdate}
\lambda_{l}^{[i \leftarrow j]} = 2\, \tanh^{-1} \left( {\prod\limits_{k \in \set{N}(j)\setminus i} {\tanh \frac{{\lambda_{l - 1}^{[k \to j]}}}{2}} } \right),
\end{equation}
which computes the message to be sent from the $j$th check node to the adjacent $i$th variable node, during the first-half of the $l$th iteration.
We update variable nodes in the usual way during the second-half of every iteration.
In \eqref{CNupdate}, we use $\set{N}(j)\setminus i$ to indicate all variable nodes adjacent to check node $j$, excluding variable node $i$.
We denote the sum of the decoder's intrinsic and extrinsic LLRs at the $i$th variable node at the completion of the $l$th iteration by $\tilde\lambda_l^{[i]}$.

\subsection{SPA Decoder without Saturation}
\label{ss-SPAwo}
Direct implementation of (\ref{CNupdate}) yields numerical problems at high LLRs.
As specified in IEEE 754 \cite{754}, double-precision floating-point (\textit{i.e.}, $64$-bit) computer computations maintain $53$-bits of precision,
while the remaining bits are for the sign and exponent.
Thus, such a computer implementation results in $\tanh(\lambda/2)$ being rounded to $\pm1$ for any LLR magnitude $|\lambda|>38.1230949 = 55\ln2$.

As an argument of $\pm 1$ will cause the $\tanh^{-1}$ function to overflow, protection from high magnitude
LLRs must be added to (\ref{CNupdate}) or the variable node update expression to ensure numerical integrity or an alternative
solution not using $\tanh^{-1}$ must be found.
Thus, preventing $\tanh^{-1}$ overflow by limiting LLRs will result in an upper limit on LLR magnitude.
This is what we refer to as ``saturating" in an LLR-domain SPA decoder.
Our examination of published error floor results suggests that LLR saturation is commonly employed.

Decoder hardware implementations also typically saturate the LLRs, although for different reasons.  
Hardware saturation is usually at LLR levels less than the $38.1$ mentioned above.
This leads to the situation where both floating-point simulation and hardware designs produce error floors at performance levels with minor variation.
The efforts to explore beyond these barriers have primarily appeared only in the very recent literature, including \cite{XiaojieFloor,SCH12,ButlerGC}.

The non-saturating SPA decoder simulation we have used is based on a pairwise check node reduction \cite{ChenHuSPA,AnastSPA}
which includes a small approximation \cite{Richter05}.
Its implementation in double-precision floating-point contains no numerical issues until LLR magnitudes reach approximately $1.79 \cdot 10^{308}$.
For a detailed examination of the numerical issues of the SPA decoder in several domains of floating-point computation, see \cite{ButlerGC}.

\subsection{Trapping Sets and Absorbing Sets}
\label{ss-tsas}

\begin{defn}[Richardson \cite{RichFloors}]
\label{def_trapping}
A \emph{trapping set} is an error-prone set of variable nodes $\set{T}$ that depends upon
the decoder's input space and decoding algorithm.
Let $\tilde\lambda_l^{[i]}(\vect{r})$ denote the $i${th} soft output symbol at the $l${th} iteration given the decoder input vector $\vect{r}$.
We say that symbol $i$ is \emph{eventually correct} if there exists $L$ such that, for all $l \ge L, \tilde\lambda_l^{[i]}(\vect{r})$ is correct.
If the set of symbols that is not eventually correct is not the empty set, we call it a trapping set, $\set{T(\vect{r})}$.
It is called an ($a, b$) trapping set if the set has $a=|\set{T}|$ variable nodes and the induced subgraph $B_{\set{T}}$ contains exactly $b$
check nodes of odd degree.
\end{defn}

From here on this \paperchapt/ will use the term \emph{trapping set}, unqualified, to mean the trapping sets defined above
with respect to the AWGN channel decoded by an LLR-domain SPA decoder with saturating LLRs.
Note, that while near-codewords are distinct from codewords, the trapping set definition includes codewords.

\begin{figure*}[!t]
\centerline{
\subfloat[$a=3$, $b=1$]{\includegraphics[height=1.40in]{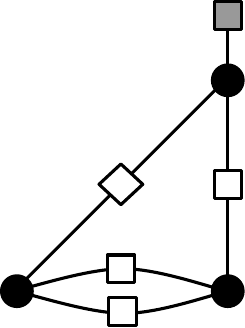}
\label{fig_31}}
\hfil
\subfloat[$a=4$, $b=2$]{\includegraphics[width=1.5in]{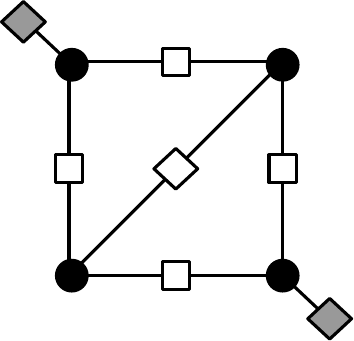}
\label{fig_42}}
\hfil
\subfloat[$a=5$, $b=3$]{\includegraphics[width=1.5in]{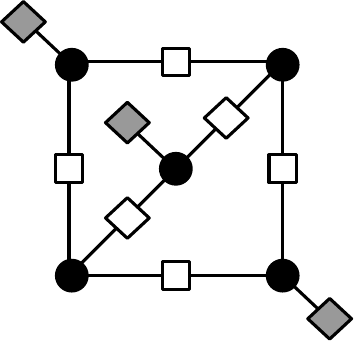}
\label{fig_53}}
\hfil
\subfloat[$a=4$, $b=4$]{\includegraphics[width=1.5in]{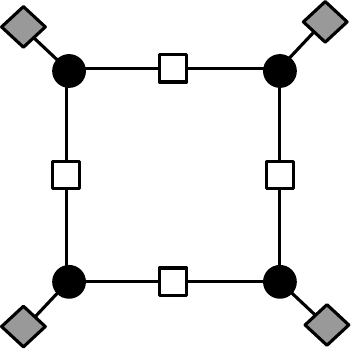}
\label{fig_44}}
}
\caption{Four sample elementary Tanner subgraphs of a code with $\dv = 3$. Odd-degree check nodes are shown shaded.}
\label{fig_trap}
\end{figure*}

\begin{defn}
A Tanner subgraph or trapping set is called \emph{elementary} if all the check nodes are of degree one or two \cite{Laendner09}.
\end{defn}

Elementary trapping sets are simple enough to be modeled with linear systems
and also account for most of the error floors seen in practice.
In \cite{Han09,RichFloors,Laendner09,Cole,ZhangRyanFloor,Milenk07}, the authors observe 
that the majority of trapping sets contributing to the error floors of belief propagation decoders are elementary trapping sets.

\begin{exmp}
Four sample elementary Tanner subgraphs are shown in Fig.~\ref{fig_trap} for a code of variable-degree three.
These are all trapping sets if they satisfy the ``not eventually correct" condition of Definition~\ref{def_trapping}.
\end{exmp}

We will define absorbing sets as they dominate the error floor of SPA decoders with saturating LLRs.
Absorbing sets are a particular type of near-codeword or trapping set.
The fully absorbing definition adds conditions beyond the subgraph itself to the remaining Tanner graph.

\begin{defn}[Dolecek \textit{et al.} \cite{DolecekTIT}]
\label{def_absorbing}
Let $\set{S}$ be a subset of the variable nodes $V$ of the Tanner graph $B=(V,C,E)$
and let $\set{S}$ induce a subgraph having a set of odd-degree check nodes $\set{N}_\mathrm{o}(\set{S})$ and a set of even-degree check nodes $\set{N}_\mathrm{e}(\set{S})$.
If $\set{S}$ has cardinality $a$ and induces a subgraph with $|\set{N}_\mathrm{o}(\set{S})|=b$, in which each node in $\set{S}$ has strictly fewer neighbors in $\set{N}_\mathrm{o}(\set{S})$ than neighbors in $\set{N}_\mathrm{e}(\set{S})$, then we say that $\set{S}$ is an $(a,b)$ \emph{absorbing set}.
If, in addition, each node in $V$ has strictly fewer neighbors in $\set{N}_\mathrm{o}(\set{S})$ than neighbors in the set $C \setminus \set{N}_\mathrm{o}(\set{S})$, then we say that $\set{S}$ is an $(a,b)$ \emph{fully absorbing set}.
\end{defn}

Fully absorbing sets are stable structures during decoding using the bit-flipping algorithm \cite{DolecekTIT}.
This results from every variable node of the set neighboring strictly more checks which reinforce the incorrect value than
checks working to correct the variable node.

\section{State-Space Model} 
\label{sect-ss}
In this section we refine and justify the existing linear system model which closely approximates the non-linear message passing of the LLR-domain SPA decoder
described earlier, as applied to an elementary trapping set.
Since this is a linear block code, with no loss of generality, we will frequently assume for convenience that the all-zero codeword has been sent.

The state-space model was introduced in \cite{SunPhD,SunAller} to analyze elementary trapping sets.
Identifying and modeling trapping sets are of interest so as to explain the observed phenomenon
of the trapping sets' variable nodes being decoded incorrectly, while the variable nodes outside of the set are eventually
decoded correctly by an LLR-domain SPA decoder.

\begin{ass}
\label{ass_elem}
All trapping sets of interest are elementary and contain more than one variable node.
We do not consider subgraphs with a single variable node as they would not have states in the state-space model.
For trapping-like behavior of single degree-$1$ variable nodes, see \cite[p.~24]{Hamk}. 
\end{ass}

First, we review the failure state of an elementary trapping set.
Let the variable nodes of the trapping set $\set{S} \subset V$ induce the Tanner subgraph $B_\set{S}$.
In later iterations the condition is reached where the variable nodes $\set{S}$ within the trapping set
are in error and those variable nodes outside the trapping set (\textit{i.e.}, $V \setminus \set{S}$) are correct.
In this condition, the subgraph's check nodes of degree one are unsatisfied and those of degree two are satisfied.

The vector of messages from the degree-one (\textit{i.e.}, unsatisfied) check nodes $\vlambda_l^{(ex)}$ at iteration $l$
are taken to be stochastic system inputs to the state-space model and are therefore modeled separately.
We will use density evolution and simulation techniques, described in later sections, to model this input vector.
The other system input vector is the intrinsic information $\vlambda$ provided by the
channel to be used in variable node updates at every iteration.
For the AWGN channel model, each element of $\vlambda$ is an independent and identically distributed (i.i.d.) Gaussian random variable.
Both of these system inputs will be treated as column vectors of LLRs; the vector $\vlambda$ has $a$ entries and
the vector $\vlambda_l^{(ex)}$ has $b$ entries for an $(a,b)$ trapping set.

\subsection{Check Node Gain Model}
\label{ss-cngm}
\begin{figure}
\centering
\includegraphics[width=2in]{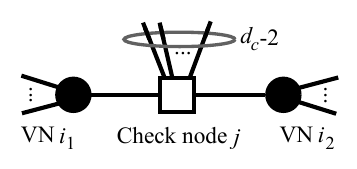}
\caption{Illustration of a check node of degree two in the trapping set and degree $\dc$ in the Tanner graph.}
\label{figcheckdeg2}
\end{figure}

Sun's linear model included the asymptotic approximation that every degree-two (\textit{i.e.}, satisfied) check node output message is equal to the 
input message at the other edge \cite{SunPhD}. 
This is based on the conditions of correct variable nodes and very high LLR values outside of the trapping set.
Schlegel and Zhang introduced more accuracy to this model by applying a
multiplicative gain $g_l$ at the degree-two check nodes, where $0 < g_l \le 1$ \cite{SCH10}.
This gain models the effect of the $\dc - 2$ external variable nodes lowering the magnitude of LLR messages as
they pass through the degree-two check nodes. 
This is illustrated in Fig.~\ref{figcheckdeg2}.
This approach adds accuracy to modeling the early iterations, and after several iterations these gains approach $1$.

\begin{ass}
\label{ass_cindep}
We will assume that the external inputs to the trapping set's degree-two check nodes are i.i.d.
This is generally a reasonable assumption to make as the gain computations are most significant at early iterations.
By the iteration count in which significant message correlation is present on the check nodes' inputs, the gains are approximately unity.
\end{ass}

Schlegel and Zhang compute the mean gain $\bar{g}_l$ corresponding to each iteration count $l\ge 1$ \cite{SCH10}.
Thus, the gain model turns the state updates into a time-varying process.
We have tried introducing a gain variance to the model, but found the effect to be very small,
so we present their mean gain model\footnote{The following includes a correction to the gain expressions of \cite{SCH10} related to the iteration count dependencies}, for which the gain of the $j$th check node during the $l$th iteration is
\begin{equation}
\label{gaindef}
g_l^{[j]} \triangleq \prod\limits_{k\in\set{N}(j) \setminus \{i_1,i_2\}} \tanh \left( \frac{{\lambda_{l - 1}^{[k \to j]}}}{2} \right).
\end{equation}
Then the expected value of the check node's gain over all realizations of the channel noise vector $\vect{n}$ 
and all check nodes $c_j$, such that $d(c_j)=\dc$,
is $\bar{g}_l(\dc) = \mathbb{E}_{\vect{n},j}\bigl[ g_l^{[j]} \bigr]$.
Using Assumption~\ref{ass_cindep}, one forms the following simplification with respect to $\bar{g}_l(\dc)$
\begin{equation}
\label{gmean2}
\bar{g}_l(\dc) = \mathbb{E}_{\vect{n},j,k} \left[ \tanh \left( \frac{{\lambda_{l - 1}^{[k \to j]}}}{2} \right) \right]^{\dc-2},
\end{equation}
over all $j$ such that $d(c_j)=\dc$ and all $k\in \set{N}(j)$.

The application of gain at the check node may be justified using a Taylor series expansion of (\ref{CNupdate}) that
begins with letting
\begin{equation}
\label{gmodel4}
\lambda^\mathrm{out}_l \triangleq f(\lambda) = 2\, \tanh^{-1} \left[ \tanh \left( \frac{\lambda}{2} \right) g_l \right],
\end{equation}
where $g_l$ is the check node's gain (\ref{gaindef}), 
$\lambda$ is the check node's input LLR from the trapping set,
and $\lambda^\mathrm{out}_l$ is the check node's output LLR on the other edge within the trapping set.
Noting that the first two partial derivatives of (\ref{gmodel4}) evaluated at $\lambda=0$ are
\begin{equation*}
\label{gmodel5}
{\left. {\frac{\partial     f(\lambda)}{{\partial \lambda    }}} \right|_{\lambda = 0}} = g \quad\mbox{and}\quad
{\left. {\frac{\partial^2 f(\lambda)}{{\partial \lambda^2}}} \right|_{\lambda = 0}} = 0,
\end{equation*}
one produces the truncated series approximation $\lambda_\mathrm{out} \approx g \lambda$.
The approximation is good for $|\lambda| \le 2 \tanh^{-1} (g)$. 
The exact expression for $\lambda_\mathrm{out}$ (\ref{gmodel4}) saturates at an output of $\pm 2 \tanh^{-1} (g)$, 
while the approximation is linear in $\lambda$.
Since we expect the LLRs outside of the trapping set to grow quite fast, we believe this to be an adequate approximation.

We now extend the check node gain model to include inversions through the degree-two check nodes caused by
erroneous messages from outside of the trapping set. 
This occurs primarily during early iterations and is illustrated in Fig.~\ref{figcheckdeg2}.
When an odd number of the $\dc-2$ messages from outside of the trapping set into a degree-two check node within the trapping set are erroneous,
then that check node will invert the polarity of the LLR message sent from variable node $i_1$ to $i_2$, as well as the message sent in the opposite direction.

We will present a simplification of the method used by Schlegel and Zhang to account for these inversions \cite{SCH10}.
Moreover, they present a small rise in their predicted error floor instead of lowering it as we find with our technique.
Schlegel and Zhang injected a stochastic cancellation signal into the state update equations, while we
choose to modify the mean check node gains.

During the first iteration, messages from variable nodes may contain inversions due to channel errors.
The probability that a specific input to the check node contains an inversion during iteration $l=1$ 
is just the uncoded symbol error rate 
\begin{equation}
\label{cinv1}
P_{\mathrm{e},1} = \QF {\left( \sqrt{ \frac{ 2 R \Eb}{N_0} } \right)}.
\end{equation}
Thus, utilizing Assumption~\ref{ass_cindep}, the probability of a polarity inversion in a specific check node is given at iteration $l$ by
\begin{equation}
\begin{split}
\label{cinv2}
P_{\mathrm{inv},l} &= \sum_{k\text{ odd}} \binom{\dc-2}{k} P_{\mathrm{e},l}^{k} (1-P_{\mathrm{e},l})^{\dc-2-k}\\
&= \frac{1-(1-2P_{\mathrm{e},l})^{\dc-2}}{2},
\end{split}
\end{equation}
which is the probability of an odd number of errors in the $\dc-2$ input messages at iteration $l$.
The final simplification in \eqref{cinv2} is from \cite[p.~38]{Gal63}.

For additional iterations we can use density evolution \cite{RUDE,RUDDE} to predict an effective $\Eb/N_0$ at the output of the variable nodes
and reuse equations (\ref{cinv1}) and (\ref{cinv2}). 
When there is a channel inversion, the output message magnitude will likely be very low, as suggested in \cite{SCH10}.
Hence, to model this we will equate random channel inversions to randomly setting the check node gain to zero with probability $P_{\mathrm{inv},l}$.
We introduce the check node's \emph{modified mean gain} as simply
\begin{equation}
\label{cinv3}
\bar{g}_l' = \bar{g}_l (1 - P_{\mathrm{inv},l}).
\end{equation}

For an LDPC code with regular check node degree $\dc$, the scalar gain $\bar{g}_l'$ of (\ref{cinv3}) may be applied at each iteration $l$ to model 
the external influence on the subgraph's degree-two check nodes.
For an LDPC code with irregular check degree, we have a few options.
If the check node degree spread is small, we may generalize (\ref{gmean2}) and (\ref{cinv3}) by also taking the expectation over 
the degree distribution of all the degree-two check nodes in the subgraph.
Alternatively, we may maintain several versions of gain $\bar{g}_l'$, one for each check-degree that appears in the subgraph.

\begin{ass}
We assume that all the degree-two check nodes in the subgraph 
may be modeled by the modified mean gain $\bar{g}_l'$.
\end{ass}

One may criticize this gain model with inversions as double-counting.
First, inversions from outside the trapping set will reduce the check node gain through \eqref{gmean2}.
Then, the same inversions reduce the check node gain again in \eqref{cinv3}.
Heuristically, one may consider this as just adding weight to the significance of these inversions.
We find it justified empirically in Section~\ref{sect-numer} and note that our earlier arguments paralleled those in \cite{SCH10}.

\subsection{Linear State-Space Model}
\label{ss-lssm}

\begin{ass}
\label{ass_vreg}
We consider only codes described by $\dv$-variable-regular Tanner graphs with $\dv \ge 3$.
For an examination of irregular codes, see \cite{SunPhD}.
\end{ass}

\begin{ass}
In justifying the model it is assumed that the messages within the trapping set under study have little impact on the rest of the Tanner graph.
This seems reasonable given that the number of unsatisfied check nodes which form the main interface between the trapping set and the rest of the Tanner graph is small.
\end{ass}

The state-space modeling equations are shown below for input column vectors $\vlambda_l^{(\mathrm{ex})}$ and $\vlambda$
and for output column vector $\tilde\vlambda_l$, whose entries are the LLR-domain soft output decisions at iteration $l$ \cite{SunPhD,SCH10}.
\begin{alignat}{2}
\notag 
\vect{x}_0 &= \matr{B} \vlambda\\
\label{SS2}
\vect{x}_l &= \bar{g}_l' \matr{A} \vect{x}_{l-1} + \matr{B} \vlambda + \matr{B}_{\mathrm{ex}} \vlambda_l^{(\mathrm{ex})} &\quad&\text{for $l\ge1$}\\
\label{SS3}
\tilde\vlambda_l &= \bar{g}_l' \matr{C} \vect{x}_{l-1} + \vlambda + \matr{D}_{\mathrm{ex}} \vlambda_l^{(\mathrm{ex})} &&\text{for $l\ge1$}
\end{alignat}
The central part of the state-space model, of course, is the updating of the state $\vect{x}_l$.
The state vector $\vect{x}_l$ represents the vector of LLR messages sent along the edges from the subgraph's variable nodes toward the degree-two check nodes.
The state is updated once per full iteration.
One may consider that $\bar{g}_l' \vect{x}_{l-1}$ represents the LLR messages after passing through the degree-two check nodes 
during the first half-iteration, and $\bar{g}_l' \matr{A} \vect{x}_{l-1}$ represents their contribution to the variable node update during the second half-iteration of iteration $l$.
Since $a \dv$ equals the number of edges of the variable-regular Tanner subgraph of $a$ variable nodes, the number of states in the model of a subgraph is $m =a \dv - b$.
For an elementary subgraph, these $m$ edges are incident on the subgraph's degree-two check nodes, forcing $m$ to be an even integer.

The $m\times a$ matrix $\matr{B}$ and the $m\times b$ matrix $\matr{B}_{\mathrm{ex}}$ are used to map $\vlambda$ and $\vlambda_l^{(\mathrm{ex})}$, respectively,
to the appropriate entries of the state vector to match the set of variable node update equations.
Since every variable node has exactly one element from $\vlambda$ participating,
$\matr{B}$ has a single $1$ in every row, and is zero otherwise.
The number of $1$ entries per row of $\matr{B}_{\mathrm{ex}}$ corresponds to the number of degree-one check nodes adjacent to the
corresponding variable node, which may be none.
Like all the matrices appearing in the state-space equations, they are $(0,1)$-matrices.

The $m\times m$ matrix $\matr{A}$ describes the dependence of the state update upon the prior state vector.
Each nonzero entry $[\matr{A}]_{ij}=1$ indicates the $j${th} edge of the prior iteration contributes to the variable node update 
computation of the $i${th} edge.
Thus, the matrix $\matr{A}^T$ may be taken as the adjacency matrix associated with a simple digraph of order $m$ which describes the state-variable update relationships.
Given our $\dv$-variable-regular codes, the row weight of $\matr{A}$ and the row weight of $\matr{B}_{\mathrm{ex}}$ will sum to $\dv-1$ for every row.

Finally, the $a\times m$ matrix $\matr{C}$ and the $a\times b$ matrix $\matr{D}_{\mathrm{ex}}$ are used to map $\vect{x}_{l-1}$ and $\vlambda_l^{(\mathrm{ex})}$ entries,
respectively, to the corresponding entry of the soft output decision vector $\tilde\vlambda_l$.
The row weight of $\matr{C}$ and the row weight of $\matr{D}_{\mathrm{ex}}$ will sum to the variable node degree $\dv$ for every row.

If there is a single degree-one check node neighboring every variable node in the subgraph, such as in the $(8,8)$ trapping set
of 802.3an, then (\ref{SS2}) degenerates to the case where $\matr{B} = \matr{B}_{\mathrm{ex}}$ and $\matr{B}$ has
regular column weight as derived in \cite{SCH10}.
Furthermore, this degenerate case produces only $\matr{A}$ matrices which have uniform row and column sums with a dominant eigenvalue $r = \dv-2$. 
Thus, our development will be significantly more general than \cite{SCH10}.

Again, note that the check node gains form our approximate model for the behavior of degree-two check nodes within the trapping set.
One drawback to any linear system approach is that saturation cannot be introduced to the state variables.
Any saturation effects must be applied to the linear system inputs.

\begin{figure*}[!t]
\centerline{
\subfloat[Tanner subgraph, $B_\set{S}$]{\includegraphics[width=1.6in]{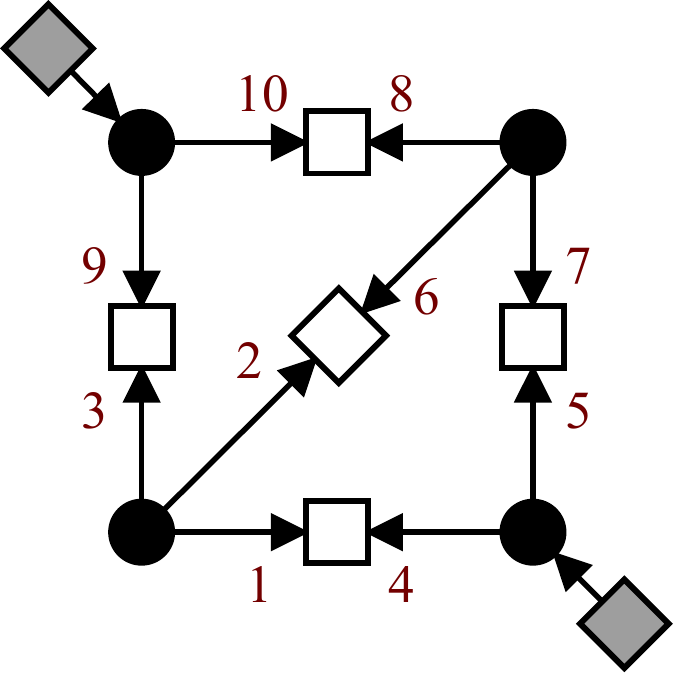}
\label{fig42trapnumB}}
\hfil
\subfloat[Multigraph, $G$]{\includegraphics[width=1.05in]{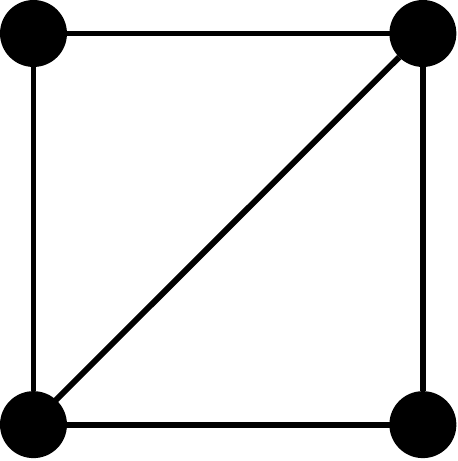}
\label{fig42trapG}}
\hfil
\subfloat[Line graph, $\mathcal{L}(G)$]{\includegraphics[width=1.05in]{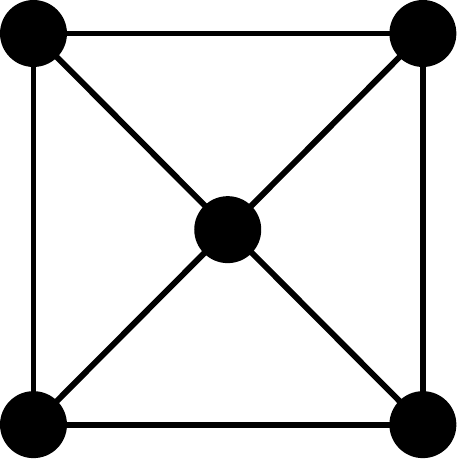}
\label{fig42linegraph}}
\hfil
\subfloat[Simple digraph, $D$]{\includegraphics[width=1.8in]{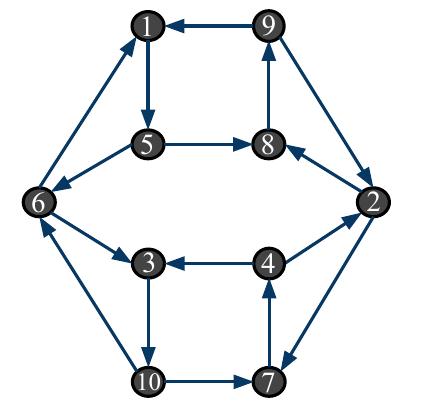}
\label{fig42digraph}}}
\caption{Several graphical descriptions of the $(4,2)$ trapping set with model's states numbered. 
Arrows have been added to the Tanner subgraph in (a) only to emphasize the direction of state variable flow.
The spectral radii of $G$, $\mathcal{L}(G)$, and $D$ are $2.5616$, $3.2316$, and $1.5214$, respectively.
}
\label{fig42graphs}
\end{figure*}

\begin{exmp}
For the $(4,2)$ trapping set with $\dv=3$, we label with integers all the edges into every degree-two check node as depicted in Fig.~\ref{fig42trapnumB}.
Now, we can set up the matrices used to make the linear system.
We have numbered the edges in Fig.~\ref{fig42trapnumB} to correspond to the order that the edges will appear in the state vector $\vect{x}_l$.
The matrix $\matr{A}$ describes the updating relationship of the state vector in a full iteration of the SPA decoder.
Since edge $1$ depends only on edges $6$ and $9$ during one iteration, we fill the first row of $\matr{A}$ to indicate such.
All the matrices associated with the trapping set of Fig.~\ref{fig42graphs} are given below.
Additionally, treating the matrix $\matr{A}^T$ as an adjacency matrix, it describes the simple digraph shown in Fig.~\ref{fig42digraph}.
\begin{equation*}
\matr{A} = \left[\begin{IEEEeqnarraybox*}[\mysmallarraydecl][c]{,c/c/c/c/c/c/c/c/c/c,}
  0&0&0&  0&  0&  1&  0&  0&  1&  0\\
  0&0&0&  1&  0&  0&  0&  0&  1&  0\\
  0&0&0&  1&  0&  1&  0&  0&  0&  0\\
  0&0&0&  0&  0&  0&  1&  0&  0&  0\\
  1&0&0&  0&  0&  0&  0&  0&  0&  0\\
  0&0&0&  0&  1&  0&  0&  0&  0&  1\\
  0&1&0&  0&  0&  0&  0&  0&  0&  1\\
  0&1&0&  0&  1&  0&  0&  0&  0&  0\\
  0&0&0&  0&  0&  0&  0&  1&  0&  0\\
  0&0&1&  0&  0&  0&  0&  0&  0&  0%
\end{IEEEeqnarraybox*}\right]
\ \ 
\matr{B} =\left[\begin{IEEEeqnarraybox*}[\mysmallarraydecl][c]{,c/c/c/c,}
   1&   0&   0&   0\\
   1&   0&   0&   0\\
   1&   0&   0&   0\\
   0&   1&   0&   0\\
   0&   1&   0&   0\\
   0&   0&   1&   0\\
   0&   0&   1&   0\\
   0&   0&   1&   0\\
   0&   0&   0&   1\\
   0&   0&   0&   1%
\end{IEEEeqnarraybox*}\right]
\ \ 
\matr{B}_{\mathrm{ex}} =\left[\begin{IEEEeqnarraybox*}[\mysmallarraydecl][c]{,c/c,}
   0&   0\\
   0&   0\\
   0&   0\\
   1&   0\\
   1&   0\\
   0&   0\\
   0&   0\\
   0&   0\\
   0&   1\\
   0&   1%
\end{IEEEeqnarraybox*}\right]
\end{equation*}
\begin{equation*}
\matr{C} =\left[\begin{IEEEeqnarraybox*}[\mysmallarraydecl][c]{,c/c/c/c/c/c/c/c/c/c,}
0&0&0&1&0&1&0&0&1&0\\
1&0&0&0&0&0&1&0&0&0\\
0&1&0&0&1&0&0&0&0&1\\
0&0&1&0&0&0&0&1&0&0%
\end{IEEEeqnarraybox*}\right]\qquad
\matr{D}_{\mathrm{ex}} =\left[\begin{IEEEeqnarraybox*}[\mysmallarraydecl][c]{,c/c,}
   0&   0\\
   1&   0\\
   0&   0\\
   0&   1%
\end{IEEEeqnarraybox*}\right]
\end{equation*}
\end{exmp}

If we remove the gains $\bar{g}_l'$ from the linear state-space equations, we have a linear time-invariant (LTI) system for which we can write a transfer function.
Let the unilateral $Z$-transform of the state $\vect{x}_l$ be denoted by $\vect{X}(z)$ and the $Z$-transform of the input vector $\vlambda^{(\mathrm{ex})}_l$ be denoted by $\vLambda_{\mathrm{ex}}(z)$, where $\vect{X}(z)=\sum_{k=0}^{\infty}\vect{x}_k z^{-k}$ \cite{Kailath}.
Note that the other input vector $\vlambda$ is not a function of iteration number.
Finally, letting the soft output vector $\tilde\vlambda_l$ of the system be denoted in the $Z$-domain by $\tilde{\vLambda}(z)$, yields
\begin{equation}
\label{LTI}
\begin{split}
\tilde{\vLambda}(z) = &\left[\matr{C} \left( z \matr{I} - \matr{A} \right)^{-1} \matr{B} +       \matr{I}\right] \frac{\vlambda}{1-z^{-1}}\\
                           + &\left[\matr{C} \left( z \matr{I} - \matr{A} \right)^{-1} \matr{B}_{\mathrm{ex}} + \matr{D}_{\mathrm{ex}}\right] \vLambda_{\mathrm{ex}}(z).
\end{split}
\end{equation}
This makes evident the importance of the eigenvalues of $\matr{A}$. As the eigenvalues are the roots of the characteristic
equation $\det(\matr{A}-\mu \matr{I})=0$, they correspond directly to the poles of the discrete-time LTI system in (\ref{LTI}).
We find that for the trapping sets addressed herein, the dominant poles are on or outside the unit circle.
Hence, the LTI system is marginally stable or (more often) unstable.

We wish to take the recursive state-update equation (\ref{SS2}) and develop it into
an expression without feedback. The first two iterations are easy to express term-by-term, as
\begin{align*}
\vect{x}_1 &= \bar{g}_1' \matr{A} \matr{B} \vlambda + \matr{B} \vlambda + \matr{B}_{\mathrm{ex}} \vlambda_1^{(\mathrm{ex})} \quad\mbox{and}\\
\begin{split}
\vect{x}_2 &= \bar{g}_1' \bar{g}_2' \matr{A}^2 \matr{B} \vlambda + \bar{g}_2' \matr{A} \matr{B} \vlambda + \matr{B} \vlambda \\
&\quad+ \bar{g}_2' \matr{A} \matr{B}_{\mathrm{ex}} \vlambda_1^{(\mathrm{ex})} + \matr{B}_{\mathrm{ex}} \vlambda_2^{(\mathrm{ex})}.
\end{split}
\end{align*}
Generalizing to an arbitrary iteration $l>0$, we have
\begin{equation}
\label{e6}
\vect{x}_l = \matr{A}^l \matr{B} \vlambda \prod_{j=1}^l \bar{g}_j' + \sum_{i=1}^{l} \matr{A}^{l-i} \left(\matr{B} \vlambda + \matr{B}_{\mathrm{ex}} \vlambda_i^{(\mathrm{ex})} \right) \prod_{j=i+1}^l \bar{g}_j'.
\end{equation}

\subsection{Graphical Assumptions and Interrelationships}
\label{ss-relat}
We have already proposed two graphical restrictions in Assumptions~\ref{ass_elem} and \ref{ass_vreg}.
This subsection notes some further assumptions we wish to place on the trapping sets and describes the
interrelationships among several useful graphical descriptions of the trapping set: the Tanner subgraph, its corresponding undirected multigraph,
the line graph of the multigraph, and the simple digraph that corresponds directly to the state update model.
Finally, we form estimates of the dominate eigenvalue for the trapping set.

\begin{lem}
\label{gmap}
There exists a bijective map between the set of $\dv$-variable-regular elementary Tanner subgraphs (up to a relabeling of the check nodes) 
and the set of multigraphs $G=(V,E)$ with vertex degrees upper bounded by $\dv$, \textit{i.e.}, $d(v_i) \le \dv \forall\: v_i \in V$.
\end{lem}
\begin{IEEEproof}
Given the elementary Tanner subgraph $B_\set{S}=(\set{S},\set{N}(\set{S}),E_\set{S})$, each variable node $v \in \set{S}$ 
becomes a vertex $v \in V$ of $G$, that is $V=\set{S}$, and the degree-one check nodes are discarded.
The check nodes of degree two become the edges of $G$, each joining a pair of vertices.
Alternatively, let parity-check submatrix $\matr{H}_\set{S}$ describe $B_\set{S}$; then $\matr{A}(G)=\matr{H}_\set{S}^T \matr{H}_\set{S} - \dv \matr{I}$.

Given the multigraph $G$, each vertex $v \in V$ of $G$ becomes a variable node $v \in \set{S}$ of $B_\set{S}$.
So, again $V=\set{S}$.
Each edge of $G$ is replaced by a degree-two check node (with an arbitrary label) and two edges in $B_\set{S}$.
Finally, degree-one check nodes are attached with single edges to variable nodes as needed until every variable node has $\dv$ neighbors.
\end{IEEEproof}

In creating the multigraph $G$, we have a more basic description of the potential elementary trapping set.
This is illustrated in creating Fig.~\ref{fig42trapG} from Fig.~\ref{fig42trapnumB}.
Either one of these graphs is sufficient to characterize the iteration-to-iteration dependencies among the state variables in the model.
The multigraph $G=(V,E)$ corresponding to $B_\set{S}$ has
\begin{equation}
\begin{split}
\label{trappingG}
\order{(G)} =&|V|= a\quad\mbox{and}\\
\size{(G)}   =&|E|= \frac{a \dv - b}{2}.
\end{split}
\end{equation}
Tanner graphs with cycles of length $4$, the smallest permitted in a Tanner graph, will map to parallel edges in $G$.
Cycles of length $k$ in $B_\set{S}$ map to cycles of length $k/2$ in $G$.
The $4$-cycles in the Tanner graph are often avoided by code designers.

\begin{exmp}
The multigraph corresponding to the Tanner subgraph of Fig.~\ref{fig_44} is the cycle graph of order $4$, $C_4$.
\end{exmp}

\begin{ass}
\label{assg1}
Tanner subgraphs of interest and their associated multigraphs are connected.
Those trapping sets described by disconnected Tanner subgraphs can instead be analyzed component by component.
The main reason for this simplification is to reduce the number of cases in which the state update expression contains a reducible $\matr{A}$ matrix. 
\end{ass}

\begin{ass}
\label{assg2}
We further assume that the variable nodes within the Tanner subgraphs contain from zero to $\dv-2$ adjacent degree-one check nodes.
Equivalently, we assume that the associated multigraphs are leafless; that is, they do not contain vertices of degree one.
For consideration of multigraphs with leaves, see Appendix~\ref{app-add}.
\end{ass}

A leaf in the multigraph maps to a variable node in the Tanner subgraph neighboring one degree-two check node and $\dv-1$ degree-one check nodes.
Thus, we require here that every variable node in the Tanner subgraph must neighbor at least two degree-two check nodes.
The multigraphs allowed by Assumptions~\ref{assg1} and \ref{assg2} will be connected, leafless graphs, containing one or more cycles.

The conditions described here, Sun termed ``simple trapping sets'' in \cite{SunPhD}.
Our Assumption~\ref{assg2} is equivalent to elementary absorbing sets for codes of variable-degree three.
However, our Assumption~\ref{assg2} is less restrictive in some ways than absorbing sets for $\dv \ge 4$ as we allow for more degree-one 
check nodes per variable node.

Finally, we describe how to create the simple digraph $D=(Z,A)$ from an undirected multigraph $G=(V,E)$ which meets Assumptions~\ref{assg1} and \ref{assg2}.
This is illustrated in creating Fig.~\ref{fig42digraph} from Fig.~\ref{fig42trapG}.
Each edge of a connected leafless multigraph $G$ corresponds to two vertices in $D$, representing the two directions of LLR message flow in 
the original Tanner graph.
That is, the number of vertices $m$ in $D$ is simply
\begin{equation*}
m=\order{(D)} = 2\:\size{(G)} = a \dv - b.
\end{equation*}
The digraph $D$ is a representation of the message updating process with respect to the output of the trapping set's variable nodes in one full iteration of the SPA decoder. 
Let edge $e_i=\{v_j,v_k\} \in E$ map to vertices $z_i, z_{i'} \in Z$, with $z_i$ representing the direction of $e_i$ flowing from $v_j$ to $v_k$
and $z_{i'}$ representing the other direction.
No arcs in $D$ join $z_i$ to $z_{i'}$.
Arcs initiating in $z_i$ are directed to vertices corresponding to other edges in $G$ flowing out of $v_k$ and
arcs terminating in $z_i$ are directed from vertices corresponding to other edges in $G$ flowing into $v_j$.
Hence,
\begin{align}
\label{dminus1a}
d^+_i &= d(v_k)-1 = d^-_{i'},\\
\label{dminus1b}
d^-_i &= d(v_j)-1 = d^+_{i'}, \quad\mbox{and}\\
\notag
\size{(D)} &= \sum_{z_i \in Z} d^-_i = \sum_{v_j \in V} d(v_j)\left[d(v_j)-1\right].
\end{align}
With respect to a Tanner subgraph that meets Assumption~\ref{assg2}, induced from a code of variable-degree three,
the size of the associated digraph simplifies to $6 a - 4 b$. 
In our $(4,2)$ example, the digraph of Fig.~\ref{fig42digraph} has order $10$ and size $16$.

We now describe a second version of the construction of the simple digraph.
From the multigraph $G$, shown in Fig.~\ref{fig42trapG}, we create the line graph $\mathcal{L}(G)$ shown in Fig.~\ref{fig42linegraph}
as described in Section~\ref{ss-ggt}.
Next, we perform a special directed lifting of $\mathcal{L}(G)$ to form the simple digraph $D$, shown in Fig.~\ref{fig42digraph}.
This lifting by a factor of two replaces each vertex with a pair of vertices and each edge with a pair of arcs, oriented in opposite directions.
Thus, if the $(i,j)$ entry of $\matr{A}(\mathcal{L}(G))$ is $w$ it is replaced with a $2\times 2$ submatrix containing $w$ ones,
where the line graph limits $w$ to the values $0$, $1$, and $2$.
When the $2\times 2$ submatrix which replaces the $(i,j)$ entry of $\matr{A}(\mathcal{L}(G))$ is determined, the selection of the corresponding $(j,i)$ submatrix 
follows a set of rules of correspondence, listed in Table~\ref{t-rules}, which ensure that the new pair of arcs are oriented in opposite directions.
Thus, the subdiagonal elements can be easily determined from the superdiagonal elements.
The exact construction of either one of these is more involved and requires a direction assignment consistency between, for example, Fig.~\ref{fig42trapnumB} and the lifting of Fig.~\ref{fig42linegraph}.

\begin{table}[!t]
\renewcommand{\arraystretch}{1.3} 
\caption{Submatrix Correspondence Rules of the Directed Lifting of the Line Graph
(for $i=j$, only the all-zero matrix is allowed)
}
\label{t-rules}
\centering
\begin{tabular}{c||c}
\hline
{\bfseries Entry $(i,j)$} & {\bfseries  Entry $(j,i)$} \\
\hline\hline
$\bigl(\begin{smallmatrix}1&0\\0&0\end{smallmatrix}\bigr)$ & $\bigl(\begin{smallmatrix}0&0\\0&1\end{smallmatrix}\bigr)$\\
$\bigl(\begin{smallmatrix}0&1\\0&0\end{smallmatrix}\bigr)$ & $\bigl(\begin{smallmatrix}0&1\\0&0\end{smallmatrix}\bigr)$\\
$\bigl(\begin{smallmatrix}0&0\\1&0\end{smallmatrix}\bigr)$ & $\bigl(\begin{smallmatrix}0&0\\1&0\end{smallmatrix}\bigr)$\\
$\bigl(\begin{smallmatrix}1&0\\0&1\end{smallmatrix}\bigr)$ & $\bigl(\begin{smallmatrix}1&0\\0&1\end{smallmatrix}\bigr)$\\
$\bigl(\begin{smallmatrix}0&1\\1&0\end{smallmatrix}\bigr)$ & $\bigl(\begin{smallmatrix}0&1\\1&0\end{smallmatrix}\bigr)$\\
\hline
\end{tabular}
\end{table}

All directed walks in $D$ will correspond to walks in $G$.
All walks in $G$ that do not backtrack will correspond to directed walks in $D$.
Backtracking is prohibited here due to the structure of SPA decoder as expressed in the message update rules, which exclude an edge's own 
incoming LLR from its outgoing LLR computation.  

\begin{figure*}[!t]
\centerline{
\subfloat[cycles and a leaf]{\includegraphics[height=0.65in]{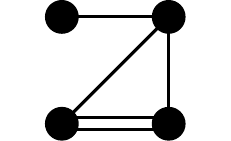}
\label{fig_inv1}}
\hfil
\subfloat[tree]{\includegraphics[height=0.65in]{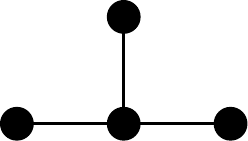}
\label{fig_inv2}}
\hfil
\subfloat[bridged cycles]{\includegraphics[height=0.65in]{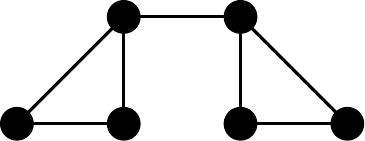}
\label{fig_val1}}}
\caption{Three sample diagrams of undirected multigraphs. 
}
\label{fig_invs}
\end{figure*}

The adjacency matrix of $D$ is $\matr{A}(D)$, which is an $m\times m$ $(0,1)$-matrix.
The matrix $\matr{A}$ used in state space model (\ref{SS2}) is equal to $\matr{A}^T(D)$ as $D$ describes the flow of messages within the model.
This connection will be convenient when discussing the properties of $\matr{A}$.

\begin{exmp}
Figs.~\ref{fig_31}, \ref{fig_42}, \ref{fig_53}, \ref{fig_44}, and \ref{fig_val1} meet Assumptions~\ref{assg1} and \ref{assg2}.
Of these, only Fig.~\ref{fig_44} has a reducible $\matr{A}(D)$ matrix and it may be analyzed in this \paperchapt/ as explained in Appendix~\ref{app-red}.
The graphs of Figs.~\ref{fig_inv1} and \ref{fig_inv2} fail Assumption~\ref{assg2} and have reducible $\matr{A}(D)$ matrices.
In the case of Fig.~\ref{fig_inv1}, the graph may first be analyzed without the leaf, and then the leaf considered as explained in Appendix~\ref{app-add}.
The case of Fig.~\ref{fig_inv2} has all zero eigenvalues as no amount of leaf removal creates a satisfactory structure.
Thus, it is not analyzed in this \paperchapt/.
\end{exmp}

We are now ready to bound the spectral radius of the graphs presented in this subsection in order to closely approximate the dominate eigenvalues of $\matr{A}(D)$ and hence $\matr{A}$.
Using \eqref{genspecG} and \eqref{trappingG}, we may bound the spectral radius of the starting multigraph $G=(V,E)$ as
\begin{equation}
\label{specG}
\dv -{b}/{a} \le \rho(G) \le \max_{v_i \in V} d(v_i),
\end{equation}
where both equalities hold for connected $G$ if and only if $G$ is regular.
The maximum value of $d(v_i)$ is limited by the parent Tanner graph to be no greater than $\dv$.

To bound the spectrum of the line graph $\mathcal{L}(G)$, we may use the relation
\begin{equation}
\label{linerel}
\rho(\mathcal{L}(G)) = \rho(\matr{A}(G) + \matr{D})-2,
\end{equation}
where $\matr{D}=\diag{(d(v_1), d(v_2), \ldots , d(v_n))}$.
This follows from the incidence matrix related expressions for each type of graph and the following fact which is easy to show:
Given any real matrix $\matr{M}$, the two products $\matr{M}^T\matr{M}$ and $\matr{M}\matr{M}^T$ have identical nonzero eigenvalues.

It is now easy to address the case when $G$ is regular of degree $k$.
Then, $\rho(G)=k$ and $k=\dv -{b}/{a}$, where $b/a$ must be an integer.
Additionally, \eqref{linerel} implies that $\rho(\mathcal{L}(G))=2k-2$ for the line graph which is itself regular\cite[p.~36]{BrualdiRyser}.
Furthermore, we find that the corresponding digraph $D$ is regular with indegree $k-1$ and outdegree $k-1$, implying that its spectral radius must be 
$\rho(D)=k-1=\dv -1 - {b}/{a}$.
This value corresponds to Schlegel and Zhang's approximation to the spectral radius of any absorbing set \cite{SCH10} which is 
\begin{equation}
\label{specSchlegel}
\rho(D)\approx \dv -1 -{b}/{a}.
\end{equation}
Alternatively, one can view \eqref{specSchlegel} as arising from the lower bound of \eqref{specG}, which is fairly tight, 
and the fact that $\rho(D) = \rho(G) - \Delta$, where $\Delta$ is typically $1$ or a little greater.  
This difference between $\rho(D)$ and $\rho(G)$ stems from the vertex-degree relationships in \eqref{dminus1a} and \eqref{dminus1b}.
We empirically examine the accuracy of \eqref{specSchlegel} in Appendix~\ref{app-tab}.

We can derive an alternative approximation based on bounding $\rho(\mathcal{L}(G))$.
From \eqref{orderLG} and \eqref{trappingG}, we find the order of $\mathcal{L}(G)$ to be $(a\dv -{b})/2$.
For the following few calculations, we assume that typically no more than one unsatisfied check node is adjacent to any variable node.
This is always the case for absorbing sets with $\dv=3$ or $4$, and likely mostly true in many other cases.
Thus, we have that $G$ contains $b$ nodes of degree $\dv-1$ and $a-b$ nodes of degree $\dv$. 
This yields the approximation to \eqref{sizeLG} that
\begin{equation*}
\size{(\mathcal{L}(G))} \approx (\dv - 1) (a \dv - 2b)/2
\end{equation*}
and the approximate bound to the spectral radius of $\mathcal{L}(G)$
\begin{equation*}
\rho(\mathcal{L}(G)) \gtrapprox 2\left( \dv -1 -\frac{b-b/\dv}{a-b/\dv} \right).
\end{equation*}
Finally, since the directional lifting technique of creating $\matr{A}(D)$ from $\matr{A}(\mathcal{L}(G))$ roughly halves the weight of each row and column, we form the new approximation
\begin{equation}
\label{specButler}
\rho(D) \approx \dv -1 -\frac{b-b/\dv}{a-b/\dv},
\end{equation}
which is also examined in Appendix~\ref{app-tab}.
Note that when $b/a$ equals $0$ or $1$ the two approximations, \eqref{specSchlegel} and \eqref{specButler}, are equal.

\section{Dominant Eigenvalues and Probability Model} 
\label{sect-domprob}
This section will simplify the linear equation used to update the states developed in the prior section to the point that we can
easily apply a probability model and predict failure rates for specific trapping sets.

\subsection{Utilizing Dominant Eigenvalues}
\label{ss-ude}
For analysis of the model, we need to simplify the powers of the matrix $\matr{A}$ that appear in (\ref{e6}).
This section largely generalizes the development by Schlegel and Zhang \cite{SCH10} with a focus on the dominant eigenvalues of the nonnegative $m\times m$ matrix $\matr{A}$ and its left eigenvectors.
This approach will avoid using the assumption made by Sun, which was that $\matr{A}$ is diagonalizable \cite{SunPhD}.
In our extensive search of potential trapping sets, shown in Appendix~\ref{app-tab}, we found a small fraction of the trapping sets have $\matr{A}$ matrices which 
cannot be diagonalized. 

Let $\mu_k \in \C$ be an eigenvalue of matrix $\matr{A}$, and let $\vect{w}_k$ be the left eigenvector associated
with $\mu_k$, such that $\vect{w}_k^* \matr{A} = \mu_k \vect{w}_k^*$,
where $\vect{w}_k^*$ is the conjugate transpose of column vector $\vect{w}_k$.
Then, by induction, for any positive integer $i$, 
\begin{equation}
\label{e7}
\vect{w}_k^* \matr{A}^i = \mu_k^i \vect{w}_k^*.
\end{equation}

Left-multiplying (\ref{e6}) by $\vect{w}_k^*$ and using (\ref{e7}) to simplify, we derive the scalar quantity
\begin{equation}
\begin{split}
\label{e8}
\vect{w}_k^* \vect{x}_l &= \mu_k^l \vect{w}_k^* \matr{B} \vlambda \prod_{j=1}^l \bar{g}_j' \\
&+ \sum_{i=1}^{l} \mu_k^{l-i} \vect{w}_k^* \left(\matr{B} \vlambda + \matr{B}_{\mathrm{ex}} \vlambda_i^{(\mathrm{ex})} \right) \prod_{j=i+1}^l \bar{g}_j'.
\end{split}
\end{equation}

We now shift to using a specific left eigenvector of the nonnegative matrix $\matr{A}$, 
the left eigenvector $\vect{w}_1$ associated with the eigenvalue of maximum modulus $r$.
Dividing the $m\times m$ nonnegative matrix $\matr{A}$ into the following two cases,
the theory of nonnegative matrices allows us to make certain statements regarding $\vect{w}_1$ and $r$ \cite{Horn,Meyer,Minc,Seneta}:
\begin{enumerate} 
\item Let the nonnegative matrix $\matr{A}$ be {irreducible}.
There is a simple eigenvalue $r$, such that $r = \rho(\matr{A})$ and $r>0$.
The associated left eigenvector $\vect{w}_1$ of $r$ is positive.
There are no other nonnegative left eigenvectors of $\matr{A}$, except for positive multiples of $\vect{w}_1$.
\item Let the nonnegative matrix $\matr{A}$ be reducible.
Letting $\matr{P}$ be an $m\times m$ permutation matrix, the matrix $\matr{A}$ may be symmetrically permuted to $\matr{A}'=\matr{P}\matr{A} \matr{P}^T$,
where $\matr{A}'$ is in block upper triangular form.
The block diagonal submatrices $\matr{A}_i'$ are either irreducible square matrices or the 1-by-1 zero matrix.
The spectrum of $\matr{A}$ is thus the union of the individual spectra $\sigma(\matr{A}_i')$.
Appendix~\ref{app-red} explains that the reducible cases allowed by Assumptions~\ref{assg1} and \ref{assg2} have all their eigenvalues located on the unit circle, 
each with multiplicity two.
For these cases, we can associate the positive left eigenvector $\vect{w}_1 = [1,1,\ldots 1]^T$ with the eigenvalue $r=1$.
Of course there exist other nonnegative eigenvectors.
\end{enumerate}
Thus, given our prior assumptions, we may associate a positive left eigenvector $\vect{w}_1$ with the real eigenvalue $r$, and
this $r$ is real with magnitude greater than or equal to that of all other eigenvalues.

Generalizing the $\beta$ that appeared in \cite{SCH10}, we define the error indicator $\beta_l \triangleq \vect{w}_1^T \vect{x}_l$.
We define $\beta_l$ this way to create a scalar indicator of trapping set error.
Consider $\beta_l$ as the scaled projection of the state vector $\vect{x}_l$ onto the positive vector $\vect{w}_1$.
Since the state vector $\vect{x}_l$ contains the internal messages of the trapping set,
the projection onto a positive vector indicates the trapping set's messages are generally 
either in the positive (correct) direction or negative (erroneous) direction.

Thus $\beta_l$ is only approximately indicative of the variable nodes' decisions.
Especially in the case where the system is inherently unstable, $r>1$, the state variables will tend to all move toward positive or negative infinity
and dominate the soft output expression (\ref{SS3}).
The domination of (\ref{SS3}) by $\vect{x}_{l-1}$ is obvious in the case that the $\vlambda_l^{(\mathrm{ex})}$ entries are saturated to finite values.
Even when this is not the case, the $\vect{x}_{l-1}$ terms still dominate the soft output expression (\ref{SS3}) 
as the state update equation (\ref{SS2}) is much like (\ref{SS3}), but (\ref{SS3}) puts even more weight on the states.

\begin{exmp}
We continue the on-going example.
The nonnegative matrices $\matr{A}$ describing the state update operations of Figs.~\ref{fig_31} and \ref{fig_42} are both primitive, 
with $r = 1.6956$ and $1.5214$, respectively.
The nonnegative matrix $\matr{A}$ describing Fig.~\ref{fig_53} is imprimitive with $h=4$ and $r = \sqrt{2}$.
This means the four eigenvalues on the spectral circle are $\pm \sqrt{2}$ and $\pm \sqrt{2}i$.
One may find that $h=4$ by visual inspection of Fig.~\ref{fig_53}, noting that every cycle length (measured in full iterations) is divisible by four.
The matrix $\matr{A}$ describing the state update of Fig.~\ref{fig_44} is reducible with $r = 1$.
\end{exmp}

Our error indicator expression simplifies if we rescale $\beta_l$ by a positive constant
to $\beta^{'}_l \triangleq {\beta_l}/\left( {r^l \prod_{j=1}^l \bar{g}_j'} \right)$, so 
\begin{equation}
\label{beta2}
{\beta_l}' = \vect{ w}_1^T \matr{B} \vlambda + \sum_{i=1}^{l} \frac{ \vect{ w}_1^T \left( \matr{B} \vlambda + \matr{B}_{\mathrm{ex}} \vlambda_i^{(\mathrm{ex})} \right)}{ r^{i} \prod_{j=1}^i \bar{g}_j'}.
\end{equation}
This expression is similar to, but more general than (1) in \cite{SCH10}, which was derived for a specific degenerate trapping set as previously described.

\subsection{Probability of Error Model}
\label{ss-poem}
The expression for $\beta^{'}_l$ in (\ref{beta2}) is a linear combination of stochastic vectors $\vlambda$ and $\vlambda_l^{(\mathrm{ex})}$.
Since elements of $\vlambda$ are i.i.d. Gaussian, linear operations on $\vlambda$ will produce a Gaussian distribution.
Several authors \cite{SCH10,RUDE} have used the approximation that the check node output LLRs, such as $\vlambda_l^{(\mathrm{ex})}$, are Gaussian, too.
The central limit theorem implies that the distribution of a linear combination of several independent check node output messages with the elements of $\vlambda$ will be nearly Gaussian, even if the check node output LLRs are only approximately Gaussian.
\begin{ass}
We assume that $\beta^{'}_l$ has a Gaussian distribution.
\end{ass}
Under this assumption, the probability of the failure event, $\xi(\set{S})$, corresponding to trapping set $\set{S}$, at iteration $l$, is then simply
\begin{equation}
\label{Qfun}
\Pr\left\{ \xi (\set{S}) \right\} = \Pr\left\{ \beta^{'}_l<0 \right\} = \QF{\left( \frac{ \mathbb{E}[\beta^{'}_l ] }{ \sqrt{\var [\beta^{'}_l ] }} \right)}.
\end{equation}
If $\{\set{S}_i\}$ enumerates all potential trapping sets, the union bound provides an upper bound on the error floor of the block error rate 
\begin{equation}
\label{UnionFER}
\Pf \lessapprox \sum_i \Pr\left\{ \xi (\set{S}_i) \right\}.
\end{equation}
The block error rate of (\ref{UnionFER}) is also commonly called the frame error rate (FER) or codeword error rate.
Next, we wish to express the error floor as an information bit error rate (BER).
Letting $\hat{a}_i$ represent the maximum number of information bits associated with trapping set failure $\set{S}_i$
for the specific encoding technique used, we may express the BER union bound as
\begin{equation}
\label{UnionBER0}
\Pb \le \sum_i \frac{\hat{a}_i}{k} \Pr\left\{ \xi (\set{S}_i) \right\},
\end{equation}
where $k$ is the number of information bits per codeword.
When the codeword is encoded \emph{systematically} the codeword bit locations are partitioned between information bits and parity bits.
For a systematic encoding, in which the bit error positions are spread uniformly with no preference to information and parity locations,
we may state $\mathbb{E}[ \hat{a}_i ] = a_i k / n$, where $a_i$ is the number of variable nodes in trapping set $\set{S}_i$ and
$n$ is the block length of the codeword.
In this case\footnote{We offer \eqref{UnionBER} as a correction to \cite{SCH10} which uses $k$ in the denominator}, (\ref{UnionBER0}) simplifies to
\begin{equation}
\label{UnionBER}
\Pb \le \sum_i \frac{a_i}{n} \Pr\left\{ \xi (\set{S}_i) \right\}.
\end{equation}

\subsection{Codewords}
\label{ss-cw}
In the case of elementary trapping sets that are also codewords, $b=0$ and $\matr{B}_{\mathrm{ex}}$ is not used since there are no unsatisfied check nodes.
We find that the failure probability (\ref{Qfun}) of the codeword simplifies to
\begin{equation}
\label{CWQ}
\Pr\left\{ \xi (\set{S}_{i}) \right\} = \QF {\left( \sqrt{\frac{2 R \Eb}{N_0}}\: \frac{ \sum_{k=1}^{a} (\vect{ w}_1^T \matr{B})_k }
{\sqrt{ \sum_{k=1}^{a} (\vect{ w}_1^T \matr{B})_k^2} } \right)}.
\end{equation}
This reduces further as the eigensystem of $\matr{A}$ for codewords is rather simple.
Every row sum and column sum of the matrix $\matr{A}$ is $\dv-1$, so the spectral radius is $r = \dv-1$.
The positive left eigenvector associated with $r$ is proportional to the all-one vector, $\vect{ w}_1 \propto [1,1, \ldots 1]^T$.
Also, $\matr{B}$ has uniform row weight one and column weight $\dv$.
Therefore, $(\vect{ w}_1^T \matr{B})_k = \dv \:\forall\: k \in \{1, \ldots ,a\}$.
Upon recognizing that $a$ is just the Hamming weight $w_\mathrm{H}$ in the context of codewords,
(\ref{CWQ}) simplifies to the more recognizable 
\begin{equation*}
\Pr\left\{ \xi (\set{S}_{i}) \right\} = \QF {\left(\sqrt{ \frac{2 R \Eb}{ N_0} w_\mathrm{H}} \right)},
\end{equation*}
independent of iteration count $l$.

\subsection{Non-codewords}
\label{ss-ncw}
To further understand the general behavior of (\ref{Qfun}), we examine the numerator and denominator terms.
Letting $m^{(l)}_{\lambda(\mathrm{ex})}$ be the expected value of each entry of $\vlambda_l^{(\mathrm{ex})}$, we can write
the numerator of the $\QF$-function argument as
\begin{equation}
\begin{split}
\label{Qnumer}
\mathbb{E}[\beta^{'}_l ] &=m_{\lambda} \left(1+ \sum_{i=1}^{l} \frac{1}{r^i \prod_{j=1}^i \bar{g}_j'} \right) \sum_{k=1}^{a} (\vect{ w}_1^T \matr{B})_k \\
&+ \sum_{i=1}^{l} \frac{m^{(i)}_{\lambda(\mathrm{ex})}}{r^i \prod_{j=1}^i \bar{g}_j'}\sum_{k=1}^{b} (\vect{ w}_1^T \matr{B}_{\mathrm{ex}})_k.
\end{split}
\end{equation}
\begin{ass}
\label{ass_indep}
We will assume that the entries of $\vlambda$ and $\vlambda_l^{(\mathrm{ex})}$ are statistically independent of each other at a given iteration $l$
as is often done in density evolution (DE) studies.
Further, we assume that the entries of $\vlambda_l^{(\mathrm{ex})}$ are independent from iteration to iteration, as was implicitly assumed in \cite{SCH10}.
\end{ass}
Letting $\sigma^2_{l}$ be the variance of each entry of the vector $\vlambda_l^{(\mathrm{ex})}$,
we can write the squared-denom\-inator of the $\QF$-function argument as
\begin{equation}
\begin{split}
\label{Qdenom}
\var[\beta^{'}_l ] &=2m_{\lambda} \left(1+ \sum_{i=1}^{l} \frac{1}{r^i \prod_{j=1}^i \bar{g}_j'} \right)^2 \sum_{k=1}^{a} (\vect{ w}_1^T \matr{B})_k^2 \\
&+ \sum_{i=1}^{l} \frac{\sigma^{2}_{i}}{(r^i \prod_{j=1}^i \bar{g}_j')^2}\sum_{k=1}^{b} (\vect{ w}_1^T \matr{B}_{\mathrm{ex}})_k^2.
\end{split}
\end{equation}


We are interested in the behavior of (\ref{Qnumer}) and (\ref{Qdenom}) as the number of iterations goes toward infinity.
The divergence of the first series within (\ref{Qnumer}) is not of interest as its effect will be
canceled by the first series within (\ref{Qdenom}) when they are combined into (\ref{Qfun}).
We will apply the Ratio Test to the second series within (\ref{Qnumer}) to evaluate its convergence.
In general, the Ratio Test is applied to the terms of a series $a_1, \ldots,  a_i,  a_{i+1}, \ldots$ by evaluating the ratio $\rho= \lim_{i \to \infty } \left| {a_{i+1}}/{a_i} \right|$. 
Then, the series $\sum_{i=1}^\infty{a_i}$ converges absolutely if $\rho<1$ and diverges if $\rho>1$.
Now, applied to the second series within (\ref{Qnumer}),
\begin{equation}
\label{eq_lim}
\rho= \lim_{i \to \infty } \left| \frac{m^{(i+1)}_{\lambda(\mathrm{ex})}r^{i} \prod_{j=1}^i \bar{g}_j'}
{m^{(i)}_{\lambda(\mathrm{ex})} r^{i+1} \prod_{j=1}^{i+1} \bar{g}_j'} \right|  
= \lim_{i \to \infty } \left| \frac{m^{(i+1)}_{\lambda(\mathrm{ex})}}{m^{(i)}_{\lambda(\mathrm{ex})} r} \right|.
\end{equation}
The right-most expression follows by noting that the gains $\bar{g}_l'$ approach $1$ rapidly with increasing iteration count $l$.
Thus, if the mean unsatisfied-check LLR $m^{(i)}_{\lambda(\mathrm{ex})}$ dominates $r^i$ asymptotically (\textit{i.e.}, $m^{(i)}_{\lambda(\mathrm{ex})} > C r^i$ for every positive constant $C$ and sufficiently large $i$), then $\rho>1$ and (\ref{Qnumer}) will grow without bound.

If we can be assured that (\ref{Qdenom}) does not grow as fast as the square of (\ref{Qnumer}),
then the entire argument to the $\QF$-function will grow without bound, driving the
failure rate of the potential elementary trapping set toward zero.
In that case, sufficient iterations and headroom for $\vlambda_l^{(\mathrm{ex})}$ growth are the
requirements for the model to achieve as low an error floor as desired.

Other models of error floor behavior \cite{SCH10,SunPhD} have used the Gaussian approximation
to LLR densities \cite{RUDE}, which implies $\sigma^2_{l} = 2m^{(l)}_{\lambda(\mathrm{ex})}$.
Moreover, using a numerical version of DE, Fu found that $\sigma^2_{l} < 2m^{(l)}_{\lambda(\mathrm{ex})}$ as the LLRs get large \cite{Minyue}.
With such an LLR variance, \textit{i.e.}, $\sigma^2_{l} \le c\, m^{(l)}_{\lambda(\mathrm{ex})}$ for some positive $c$, we find that the
entire argument to the $\QF$-function grows without bound in the cases that satisfy $m^{(i)}_{\lambda(\mathrm{ex})} > C r^i$.
We will find in the following sections that this latter condition is true within the assumptions identified.
However, if the variance grows as the square of the mean, then the argument to the
$\QF$-function reaches a finite limit and a nonzero error floor is produced.
This is a very important issue that we will revisit later in the \paperchapt/.

\subsection{Bounds on Spectral Radius of the Matrix $\matr{A}$}
\label{ss-bsr}

We now need some bounds on $r$, the spectral radius of the $\matr{A}$ matrix, in order to compare
it with the growth rate of $m^{(i)}_{\lambda(\mathrm{ex})}$ which we will develop later.
The classic bounds of Frobenius, from Lemma~\ref{lemFrob1}, are sufficient for our needs---merely note 
the equivalence between the outdegrees of a digraph and the row-sums of the associated adjacency matrix.


\begin{thm}
\label{rdv}
Consider a variable-regular LDPC code with variable-degree $\dv \ge 3$.
Let the trapping set $\set{S}$ induce an elementary connected subgraph $B_\set{S}$ with $a \ge 2$ variable nodes and $b>0$ degree-one check nodes.
Further, let the associated undirected multigraph $G$ be leafless.
Then, the adjacency matrix $\matr{A}(D)$ of the associated simple digraph $D$ must have spectral radius $r$ such that $1\le r<\dv-1$.
\end{thm}
\begin{IEEEproof}
The proof follows from Lemma~\ref{lemFrob1} if we defer to Appendix~\ref{app-red} the equality conditions for the cases in which the matrix $\matr{A}$ is reducible.
\end{IEEEproof}


\section{Analysis Using DE without Saturation} 
\label{sect-analnosat}

This section analyzes LLR-domain SPA decoding without saturation to make ultimate error floor predictions.
First, we must apply DE to model the contributions from nodes
outside of the trapping set relevant to the behavior of our trapping set.
Then we will refer back to the divergence condition of (\ref{eq_lim}) to see if error floors are produced.

\begin{ass}
Density evolution (DE) assumes that the Tanner graph has no cycles and the code's block length is infinite.
For our purposes, this is equivalent to the assumed independence among incoming LLRs of Assumption~\ref{ass_indep}.
\end{ass}

SPA decoding of LDPC codes exhibits a threshold phenomenon as the block length tends to infinity.
The error rate drops very dramatically as the SNR exceeds the \emph{decoding threshold} which can be found using DE \cite{RUDE}.
As we are interested in the behavior of the error floor in this work, we assume that the channel SNR is always above the decoding threshold.

The application of this DE technique produces an analytical expression for the LLRs of the SPA decoder.
As DE progresses above the decoding threshold, LLR values grow fast, but the question is how fast.
In \cite{SunPhD}, Sun showed that mean check node outputs, $m_{\lambda(\mathrm{ex})}$,
grow from iteration $l-1$ to $l$ as
\begin{equation}
\label{DE0}
m_{\lambda(\mathrm{ex})}^{(l)}=(\dv -1) m_{\lambda(\mathrm{ex})}^{(l-1)} + \mbox{``some~small~value~terms."}
\end{equation}
In \cite{ButlerDE}, the present authors seek to bound the effect of Sun's ``small value terms," by developing tighter asymptotic expressions. 
In fact, we are able to develop bounds on the SNR region in which we can expect the growth rate of the mean extrinsic LLRs $m_{\lambda(\mathrm{ex})}^{(l)}$ to exceed the internal trapping set LLR growth rate.
In any case, DE analysis in the high $\Eb/N_0$ regime leads to $m_{\lambda(\mathrm{ex})}^{(l)} > C r^l$ for every positive constant $C$, any $r$, such that $1 \le r < \dv -1$, and sufficiently large $l$.

When combined with (\ref{eq_lim}) developed in Section~\ref{sect-domprob}, the growth rates of this section show that every potential (non-codeword) elementary trapping-set error can be corrected by a non-saturating LLR-domain SPA decoder after a sufficient number of iterations, provided we can rely upon DE as a model for LLR growth outside of the trapping set for variable-regular ($\dv \ge 3$) LDPC codes.

\begin{figure}
\centering
\includegraphics[width=\figwidth]{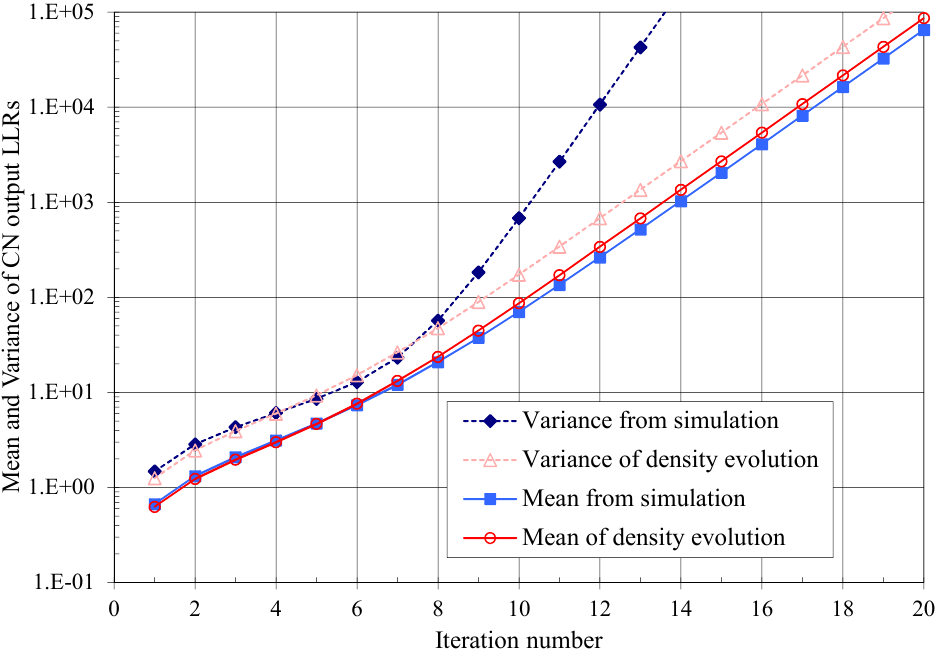}
\caption{Check node output LLR mean and variance versus iteration number for the Margulis code at $\Eb/N_0$ of $2.8$ dB.}
\label{fig_LLRG}
\end{figure}

In Fig.~\ref{fig_LLRG} we plot the mean check node output LLRs from the SPA decoder simulation of the $(2640,1320)$ Margulis code, which is $(3,6)$-regular,
and from DE of a $(3,6)$-regular code.
In this simulation the all-zero codeword is transmitted over the AWGN channel and early termination of the LLR-domain decoder is disabled.
We note that the mean LLR follows very closely the mean LLR predicted by DE.
The variance of the check-node output LLRs, however, shows a problem. 
For the first seven iterations, the variance is approximately twice the mean, as predicted by the Gaussian approximation to DE.
However, by the ninth iteration, the variance has clearly taken on the trend of the square of the mean.
As discussed in Section~\ref{sect-domprob}, this trend can generate an error floor as the mean to standard deviation
ratio of $\beta_l^{'}$ in (\ref{Qfun}) reaches a fixed value rather than growing as iterations get large, \textit{i.e.},

\begin{figure}
\centering
\includegraphics[width=\figwidth]{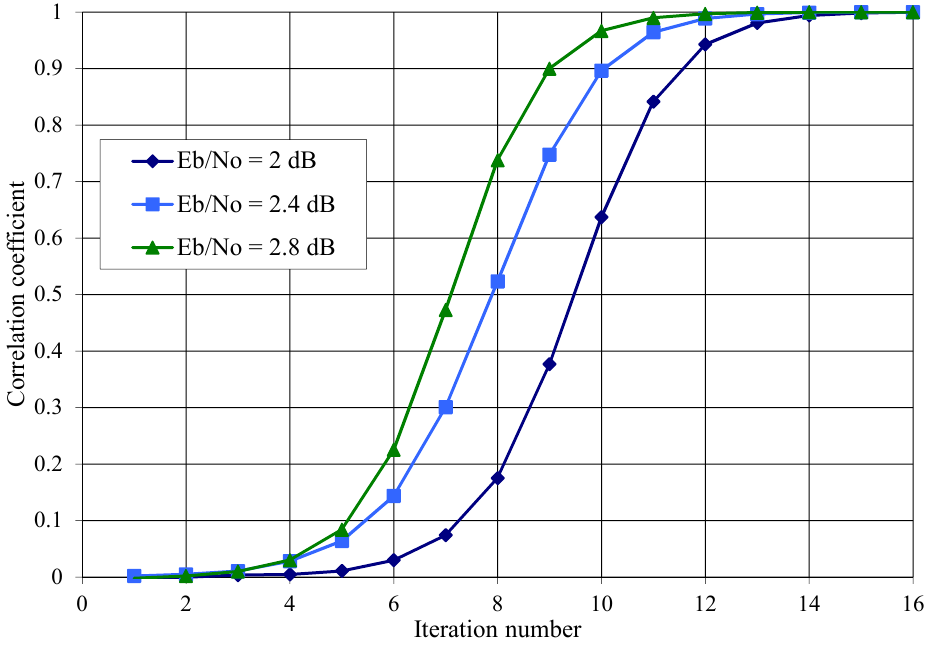}
\caption{Correlation coefficient among the four LLR check node output metrics used as linear model inputs versus iteration number for $(12,4)$ trapping sets of the Margulis code as measured in LLR-domain SPA simulation.}
\label{fig_MargCorr}
\end{figure}

\begin{equation}
\label{OEVAR}
\frac{ \mathbb{E}[\beta^{'}_l ] }{ \sqrt{\var [\beta^{'}_l ] }} = O_l(1).
\end{equation}

The cause of the variance's departure from the behavior predicted by DE is the positively correlated nature of the LLRs after several iterations.
In Fig.~\ref{fig_MargCorr} we plot the average of the off-diagonal values from the $4\times4$ matrix of correlation coefficients
of the four LLR check node output metrics used as linear model inputs versus the iteration number for the $(12,4)$ trapping sets of the Margulis code.
This shows significant positive correlation appearing in the sixth iteration and progressing rapidly.
Thus, the assumptions used to justify (\ref{Qdenom}) do not hold in this case.
The $(640,192)$ QC code, which is studied later, also shows similar variance and correlation behavior just with an earlier onset, due to its smaller girth ($6$ iterations vs. $8$).

Also exposing flaws in our probability of error model is the skewness of $\beta^{'}_l$.
We have measured its skewness in the range of $0.6$ to $1.8$ as the correlated LLRs propagate through an SPA simulation of the Margulis code.
Skewness is a measure of the asymmetry of the probability distribution of a random variable.
Of course, \eqref{Qfun} assumed a Gaussian distribution, which is symmetric about the mean and has a skewness of zero.
We consider such large skews a significant complication for this case.


The problems presented here make it impossible to accurately use the probability model that we presented beyond the seventh iteration of decoding the Margulis code without saturation.
However, in the following sections we are able to find applications of this model and use other techniques to estimate the error floor for non-saturated decoding of the Margulis code.

\section{Modifying and Applying Richardson's Semi-Analytical Tech\-nique without Saturation} 
\label{sect-margnosat}
\label{sect-margrich}
In this section, we briefly introduce Richardson's semi-analytical technique to estimate the error floor by simulating in the vicinity of trapping sets \cite{RichFloors}.
Then, we describe potential improvements to the technique for it to be effective in the case of the non-saturating decoder, and
present results for the error floor of the Margulis code.


Richardson's technique biases the noise in the direction of the trapping set (TS), much like importance sampling.
(Alternatively, one may employ importance sampling techniques to estimate the error rates of particular trapping sets, as in \cite{Cole,DolecekJSAC,SCH11}.)
Richardson effectively utilizes an orthogonal transform on the noise samples within the TS to represent its noise with a new set of basis vectors in which each dimension is still i.i.d. Gaussian noise.
This allows us to treat the Gaussian random variable $s$ in the direction of the noise separately, while we perform Monte-Carlo simulation on the other $a-1$ 
dimensions of the noise 
subvectors with variance $\sigma^2$.

We fix $s$ at several negative values and run simulations, finding the conditional TS failure probability $\Pr\left\{ {{\xi _{\set T}}|s} \right\}$ 
for each $s$ value.
Then, the overall TS failure probability $\Pr\left\{ {{\xi _{\set T}}} \right\}$ may be computed using total probability, as
\begin{equation}
\begin{split}
\label{Rich_integ}
\Pr\left\{ {{\xi _{\set T}}} \right\} &= \int {\Pr\left\{ {{\xi _{\set T}}|s} \right\}} p_S(s)\,ds. 
\end{split}
\end{equation}
Due to the high degree of overlap among TSs, when counting failures for $\Pr\left\{ {{\xi _{\set T}}|s} \right\}$, Richardson only
counts a frame as a failure if the set of symbols which are not eventually correct exactly matches the symbol locations of $\set{T}$, the specific
TS under test.

\begin{figure*}[t]
\centerline{
\subfloat[$(12,4)$]{\includegraphics[height=1.25in]{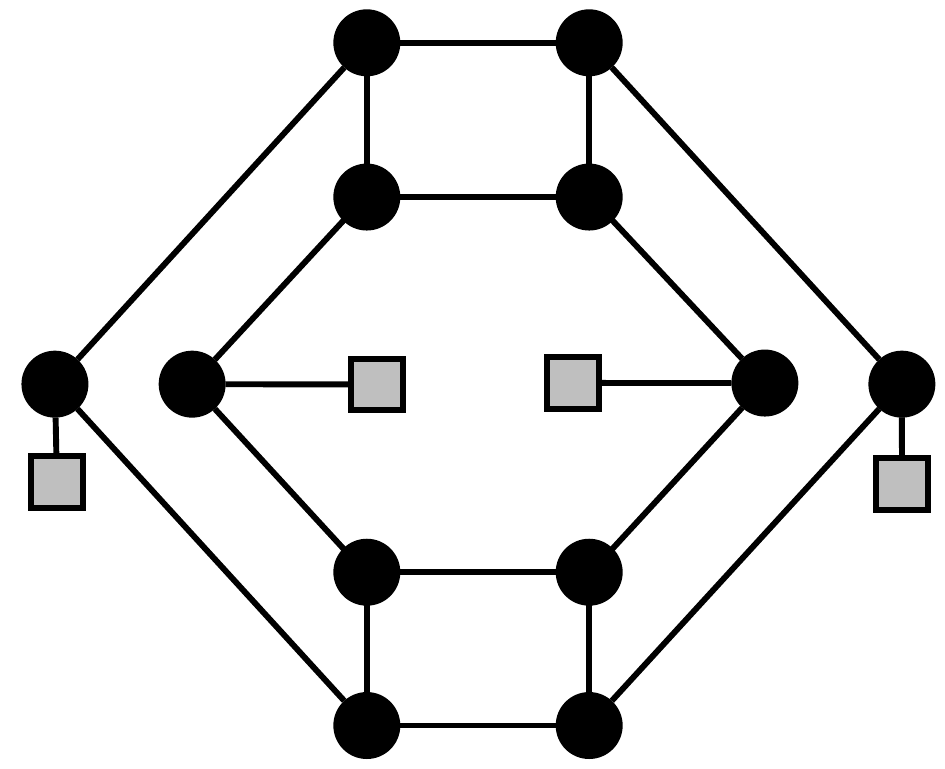}
\label{124trap}}
\hfil
\subfloat[$(14,4)$]{\includegraphics[height=1.25in]{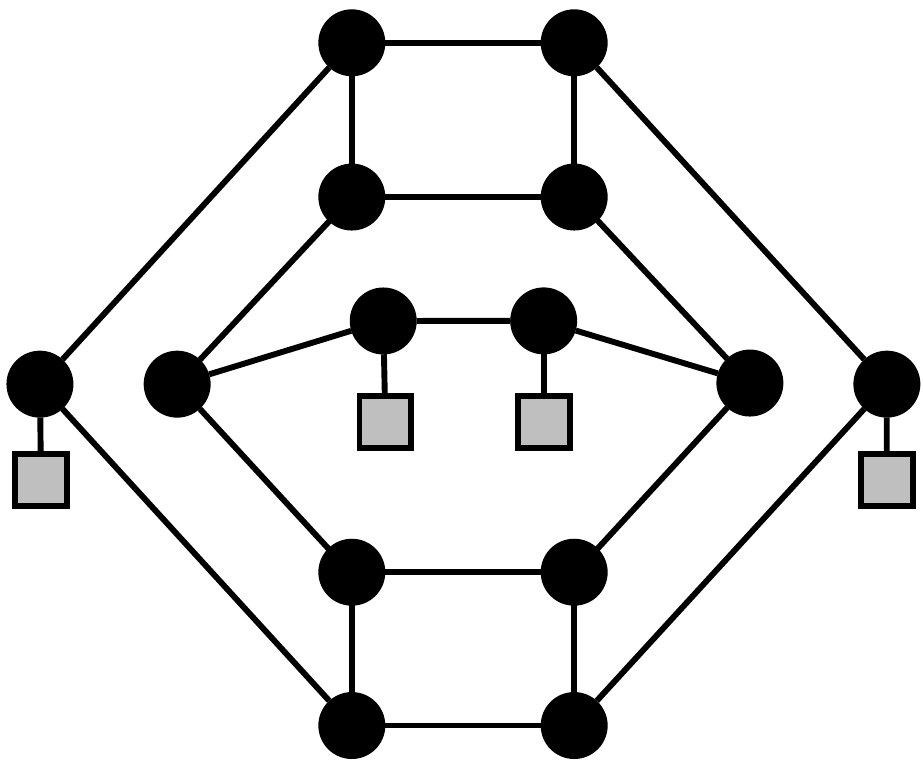}
\label{144trap}}
\hfil
\subfloat[$(18,8)$]{\includegraphics[height=1.25in]{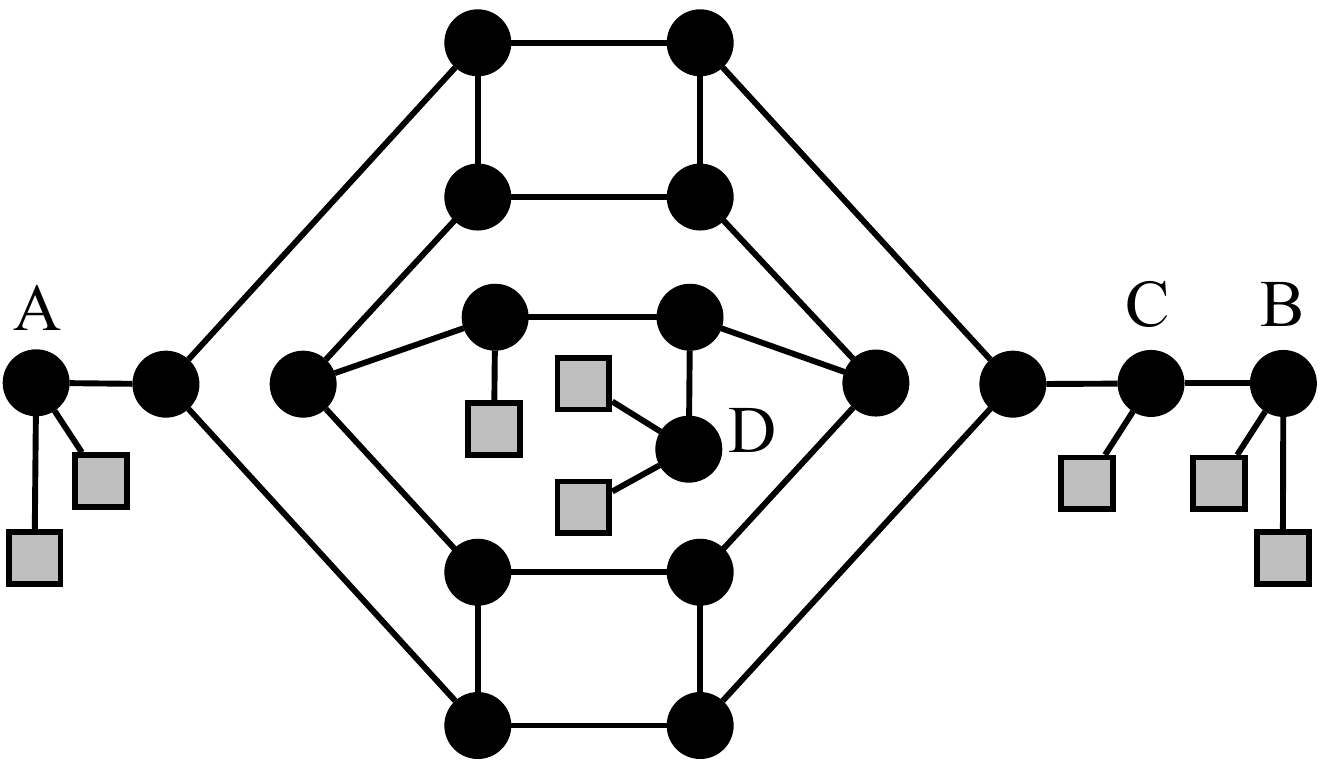}
\label{188trap}}}
\caption{Three elementary TSs of a $\dv=3$ code shown with degree-one check nodes shaded and with degree-two check nodes omitted.
The two on the left are absorbing sets and dominate the error floor of the saturated SPA decoder for the Margulis code, 
while the one on the right fails is not an absorbing set.}
\label{margtrap}
\end{figure*}

The dominant TSs  of the $(2640,1320)$ Margulis LDPC code are the $(12,4)$ and $(14,4)$ elementary TSs, shown in Figs.~\ref{124trap} and \ref{144trap} \cite{Han09,RichFloors,MKfloor}.
In the case of the $(14,4)$ TS of the Margulis code, the not-eventually-correct symbols exactly match the symbols in $\set{T}$, when 
testing the saturated decoder at a rate of about $3$ in $4$ failures or greater (at $s=-1.35$).  
Note that this rate drops substantially as $s$ approaches $-1$.
However, for a non-saturating decoder, this ratio became less than $1$ in $10^5$ (at $s=-1.35$) in our tests, indicating that the 
not-eventually-correct condition is not effective when saturation is eliminated.  
This ratio was so low when running Richardson's technique for $\set{T}$ set to a $(12,4)$ TS without saturation, that we could not measure it.
When saturation is present, failures are generally forced to occur on absorbing sets, and this is often the set under test when $s \le -1.25$.
Without saturation present, the failures occur on a wide variety of Tanner subgraphs.
Most of the failing subgraphs during non-saturating simulation are not absorbing sets and most are larger than the original $a$ variables under test.
In fact, the two smallest failing structures that we captured at high SNR were the $(14,4)$ absorbing set of Fig.~\ref{144trap} and the $(18,8)$ subgraph shown in Fig.~\ref{188trap}.
Since the set of general Tanner subgraphs within the Margulis code is vast \cite{Karimi}, 
rerunning the error floor estimate for each type of subgraph with such a low capture rate would be impractical. 
We propose two solutions to this.

The first solution would be to ignore the not-eventually-correct criteria and simply count all failures at the risk of overestimating the error floor.
Instead, we propose to estimate the error floor by introducing saturation in the final iterations of failed frames so that they may fail on absorbing sets.
For example, iterations with LLR saturation would force corrections to the labeled variable nodes of Fig.~\ref{188trap}, as the unsatisfied check nodes outnumber the 
satisfied checks, driving the failed structure to the $(14,4)$ absorbing set contained within.
Since this approach reduces the number of incorrect bits per failure, it perhaps has useful applications.
It appears effective, as we see that most failures are reduced to small absorbing sets in practice.
We prefer this strategy because we are interested in establishing a lower bound on the floor for the non-saturating case.

\begin{figure}
\centering
\includegraphics[width=\figwidth]{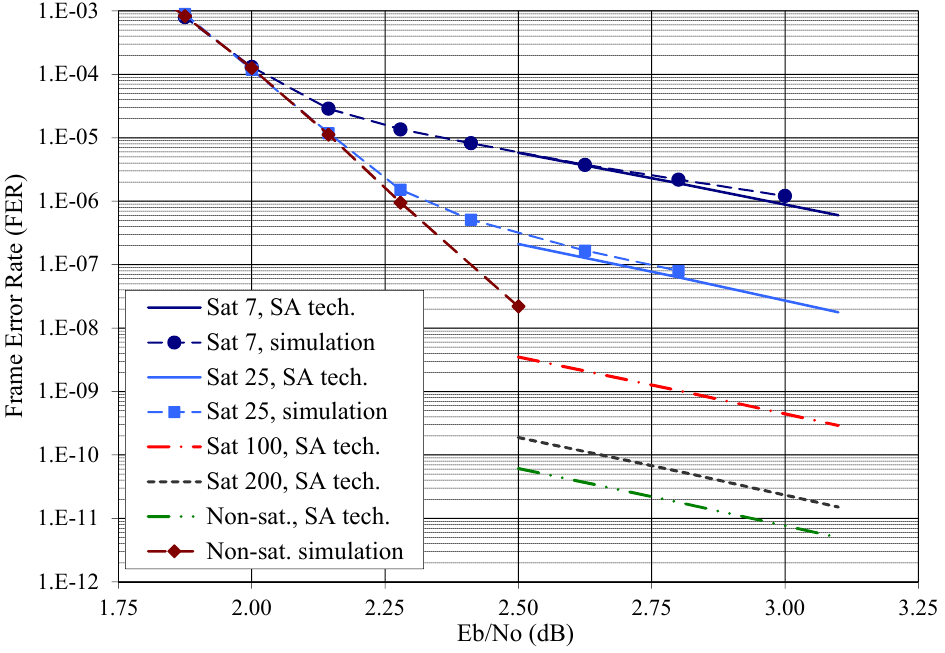}
\caption{FER vs. $\Eb/N_0$ in dB for Margulis LDPC Code in AWGN. 
The semi-analytical (SA) technique only estimates errors due to the (12,4) and (14,4) trapping sets at $\Eb/N_0 = 2.8$ dB.}
\label{fig_MFERnosat}
\end{figure}

As there may be many variable nodes to correct to get to the absorbing set contained within the failed structure we allow for $20$ iterations
of saturated SPA decoding before beginning the ``eventually correct'' logic which runs for $12$ additional iterations.
We chose to run the saturation phase at an LLR limit of $25$.
Our results are shown in Fig.~\ref{fig_MFERnosat} where all simulations were run for a maximum of $200$ iterations, except that the
non-saturating semi-analytical simulations were run for $32$ additional iterations as just described.
The semi-analytical technique was run at $\Eb/N_0 = 2.8$ dB in all cases, and was extrapolated locally as suggested in \cite{RichFloors}.
Richardson's extrapolation assumes that the conditional failure probability $\Pr\left\{ {{\xi _{\set T}}|s} \right\}$ is insensitive to SNR changes.
The extrapolations appear to produce parallel lines in the log-log plot (\textit{i.e.}, the logarithm of the error rate versus the $\Eb/N_0$ measured in decibels, which is also logarithmic) due to the character of the integration in \eqref{Rich_integ}.
The dominating part of the integrand is about $5$ standard deviations of $s$ below the zero-mean point in all the conditions studied.
Thus, small changes to the standard deviation of $s$ in \eqref{Rich_integ} produce a multiplicative effect on the estimated error rate 
regardless of the specific decoder configuration.

Fig.~\ref{fig_MFERnosat} shows an estimated error floor at $\Eb/N_0 = 2.8$ dB of $2 \cdot 10^{-11}$ FER due to these two TSs,
which is incrementally better than the $6 \cdot 10^{-11}$ achieved by limiting the LLR magnitudes to $200$ at the check node output.
In making this error floor prediction we assume that no other TSs become dominant in these conditions.

Several additional observations are worth noting.
Increasing the maximum number of iterations of the non-saturating simulation to $1000$ reduces the error floor by about a factor of $3$.
Additionally, we have applied the semi-analytical techniques of this section to estimate the combined FER contributions of the
$(12,4)$ and $(14,4)$ TSs to be $4 \cdot 10^{-17}$ FER at $\Eb/N_0 = 4$ dB.
This shows that the error floor falls significantly faster than Richardson's extrapolation predicts, 
since the conditional failure probability $\Pr\left\{ {{\xi _{\set T}}|s} \right\}$ appears to decrease significantly with increasing SNR.
From the DE expressions, at higher SNR, the beneficial LLR growth starts earlier and grows faster
giving the code beyond the TS a higher probability to correct the channel error along a TS for a non-saturating decoder.
However, an SPA decoder with limited LLR dynamic range will not leverage these effects that depend on SNR.

Table~\ref{table_ratio} shows the ratio of the error floor contributions of the $(12,4)$ TS to the contributions of the $(14,4)$ TS at $\Eb/N_0 = 2.8$ dB.
We note that when saturation limits are low the $(12,4)$ TS dominates the error floor, but when the limits are raised $(14,4)$ dominates.
There are two potential reasons for this. 
The first is that the larger maximum eigenvalue value of $(14,4)$ is more sensitive to saturation even though it is less likely to initially trigger due 
to the greater number of variable nodes involved.
Secondly, an effect we have noticed is that a growing fraction of $(12,4)$ TS failures are subsumed by the $(14,4)$ absorbing sets which contain them as LLR
limits are raised.
Additionally, we notice that the $(14,4)$ TSs dominate even more as the SNR is increased for a non-saturating decoder.

\begin{table}[!t]
\renewcommand{\arraystretch}{1.2} 
\caption{Ratio of $(12,4)$ trapping set error floor contribution to $(14,4)$ for the Margulis code at $\Eb/N_0 = 2.8$ dB
using Richardson's semi-analytical technique.
}
\label{table_ratio}
\centering
\begin{tabular}{l||r}
\hline
{\bfseries Decoder} & {\bfseries Ratio} \\
\hline\hline
Saturated SPA, at $\pm 7$ CN out    & $4.6:1$\\
Richardson's 5-bit \cite{RichFloors}     & $3.3:1$\\
Saturated SPA, at $\pm 15$ CN out   & $3.3:1$\\
Saturated SPA, at $\pm 25$ CN out   & $1.7:1$\\
Saturated SPA, at $\pm 50$ CN out   & $0.7:1$\\
Saturated SPA, at $\pm 100$ CN out  & $0.6:1$\\ 
Saturated SPA, at $\pm 200$ CN out  & $0.46:1$\\
Non-saturating SPA decoder              & $0.4:1$\\
\hline
\end{tabular}
\end{table}

Finally, we would like to know if we are seeing a  ``floor'' or not for the non-saturating SPA simulation results.
Extrapolating the FER of the waterfall region of Fig.~\ref{fig_MFERnosat} at a constant log-log slope out to $\Eb/N_0=4.0$ dB yields an FER of about $5 \cdot 10^{-20}$.
Thus, the FER contribution of the $(12,4)$ and $(14,4)$ TSs at $4 \cdot 10^{-17}$ FER 
overpowers the FER of the extrapolated waterfall region at an $\Eb/N_0$ of $4.0$ dB.  
Therefore, early evidence predicts that we will indeed see a floor, but this floor is much lower and has about twice the log-log slope versus SNR than what was previously referred to as a ``floor'' for the Margulis code.
Thus, our floor estimates are about $5$ orders of magnitude lower at $\Eb/N_0=2.8$ dB than prior SPA simulation results,
and about $8$ orders of magnitude lower at $4.0$ dB.

\section{Numerical Results with Saturation} 
\label{sect-numer}
This section readdresses the prediction model, this time for the case of saturated decoding.
We present the techniques used to produce the linear system's inputs, and then compare predicted error floors to simulation results for four LDPC codes.
Two numerical models are described for use as system inputs for the mean and variance of the unsatisfied, degree-one check nodes within the trapping set (TS).
These models also generate the mean gain $\bar{g}_l$ of the degree-two check nodes required by the model presented in Section~\ref{sect-ss}.
This completes the necessary tools to produce error floor predictions when the LLR metrics are saturated within the LLR-domain SPA decoder. 

The first technique is the numerical version of density evolution (DE) known as discretized density evolution (DDE) described in \cite{RUDDE}.
This technique utilizes a probability mass function (pmf) that closely approximates the continuous probability density function (pdf) of node input messages and output messages.
This technique is selected for the saturated case as it allows us to precisely model the behavior of LLRs as they saturate since it works directly on the pmfs.
At a particular $\Eb/N_0$ value, we capture the mean and variance of the values we require at each iteration for insertion into the model.
See Table~\ref{DEtable} for an example, which also lists the mean check node gain, which have been computed following (\ref{gmean2}).

We present our technique to contrast it with that presented in \cite{SCH10}.
Schlegel and Zhang suggest recording only the mean LLRs from a (continuous) DE analysis 
and computing the variance based on the consistent Gaussian distribution assumption \cite{RUDE}.
The consistent Gaussian approximation\footnote{The mean and variance relationship of the LLRs was printed erroneously in \cite{SCH10} as $m=2\sigma^2$}
is that $\sigma^{2}_{l} = 2\,m_{\lambda(\mathrm{ex})}^{(l)}$.
In this section, we have purposely avoided the consistency assumption since the effect of LLR saturation will void it as seen in Table~\ref{DEtable}.

The second technique is to simply run the LLR-domain SPA decoder with early termination turned off and collect the required statistics.
Configuring the AWGN channel for a particular $\Eb/N_0$ value, we capture the mean and variance of the check nodes'
output LLRs and the mean check node gain computed as in (\ref{gmean2}) at each iteration. 
For an example of the results using this technique see Table~\ref{SPAtable}.
We find $100$ frames to be sufficient, and of course very quick to produce.

The results in the two tables are similar, as expected.
Starting at iteration six or seven, the LLR variance from the SPA simulation is significantly larger than from the DDE technique, due to cycles present in the simulation of the finite length code.  This is similar to Fig.~\ref{fig_LLRG}, but under different saturation conditions.
We would expect that a code with an even a larger girth would have an even later onset of variance divergence.
Interestingly, it was found the two techniques produced nearly identical error floor predictions in spite of the differences.
We chose to present the results obtained using the SPA simulation technique in the figures that follow.

\begin{table}[!t]
\renewcommand{\arraystretch}{1.15} 
\caption{DDE numerical results by iteration for $(3,6)$-regular code at an $\Eb/N_0$ of $2.8$~dB with LLR saturation set to $\pm 25$ at check node output.
LLR pdfs discretized to pmfs with a resolution of $50/2047$.
In this case, $m_{\lambda} = 3.8109$.
}
\label{DEtable}
\centering
\begin{tabular}{c||D{.}{.}{2,3}|D{.}{.}{2,2}|D{.}{.}{1,4}}
\hline
\multicolumn{1}{c||}{\bfseries $l$} &
\multicolumn{1}{c|}{\bfseries $m_{\lambda(\mathrm{ex})}^{(l)}$} &
\multicolumn{1}{c|}{\bfseries $\sigma^{2}_{l}$} &
\multicolumn{1}{c}{\bfseries $\bar{g}_l$}\\
\hline\hline
1 &	0.669&	1.47&	0.3242\\
2 &	1.315&	2.81&	0.4898\\
3 &	2.08&	4.24&	0.6243\\
4 &	3.11&	5.94&	0.7472\\
5 &	4.66&	8.05&	0.8586\\
6 &	7.21&	10.74&	0.9446\\
7 &	11.66&	14.22&	0.9891\\
8 &	19.67&	16.59&	0.9995\\
9 &	24.91&	0.47&	1.0000\\
10 &	25.00&	0.00&	1.0000\\
\hline
\end{tabular}
\end{table}

\begin{table}[!t]
\renewcommand{\arraystretch}{1.15} 
\caption{SPA simulation results by iteration for the $(2640,1320)$ Margulis code which is $(3,6)$-regular at an $\Eb/N_0$ of $2.8$~dB with LLR saturation set to $\pm 25$ at check node output.
In this case, $m_{\lambda} = 3.8109$.
}
\label{SPAtable}
\centering
\begin{tabular}{c||D{.}{.}{2,3}|D{.}{.}{2,2}|D{.}{.}{1,4}}
\hline
\multicolumn{1}{c||}{\bfseries $l$} &
\multicolumn{1}{c|}{\bfseries $m_{\lambda(\mathrm{ex})}^{(l)}$} &
\multicolumn{1}{c|}{\bfseries $\sigma^{2}_{l}$} &
\multicolumn{1}{c}{\bfseries $\bar{g}_l$}\\
\hline\hline
1 &	0.675	& 1.49	&0.3245\\
2 &	1.324 & 2.88	&0.4897\\
3 &	2.09	& 4.36	&0.6219\\
4 &	3.14	& 6.19	&0.7425\\
5 &	4.73	& 8.83	&0.8505\\
6 &	7.37	& 13.65	&0.9338\\
7 &	12.08	& 25.82	&0.9793\\
8 &	19.13	& 33.94	&0.9956\\
9 &	23.66	& 11.86	&0.9995\\
10 &	24.83	& 1.31	&1.0000\\
\hline
\end{tabular}
\end{table}

\begin{figure}
\centering
\includegraphics[width=\figwidth]{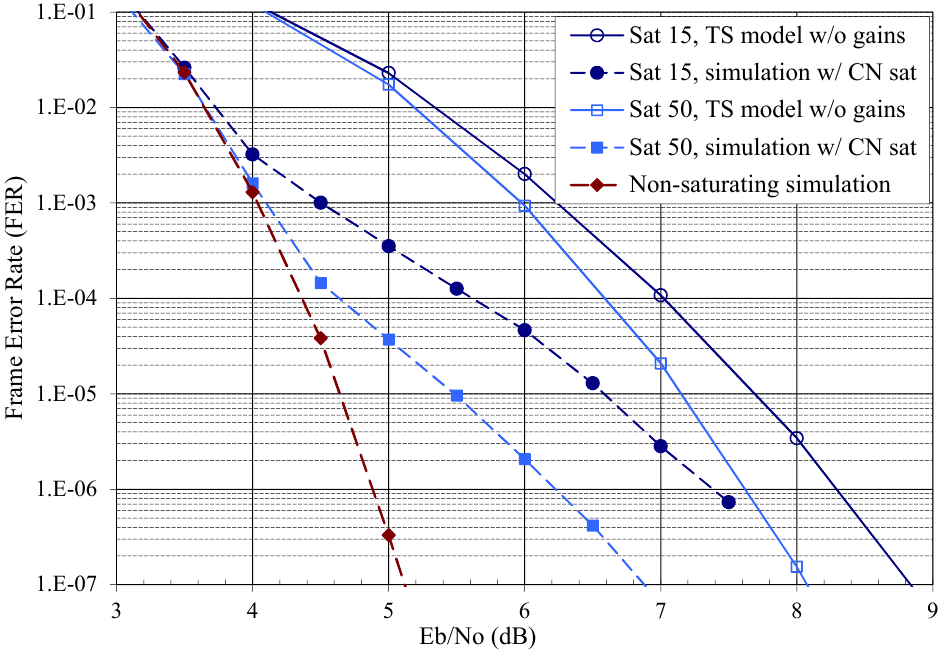}
\caption{FER vs. $\Eb/N_0$ in dB for the $(640,192)$ QC code with simulation and the TS model for predicting error floors using unit gains in the model and no check node inversions.}
\label{fig_Ryan1}
\end{figure}

\begin{figure}
\centering
\includegraphics[width=\figwidth]{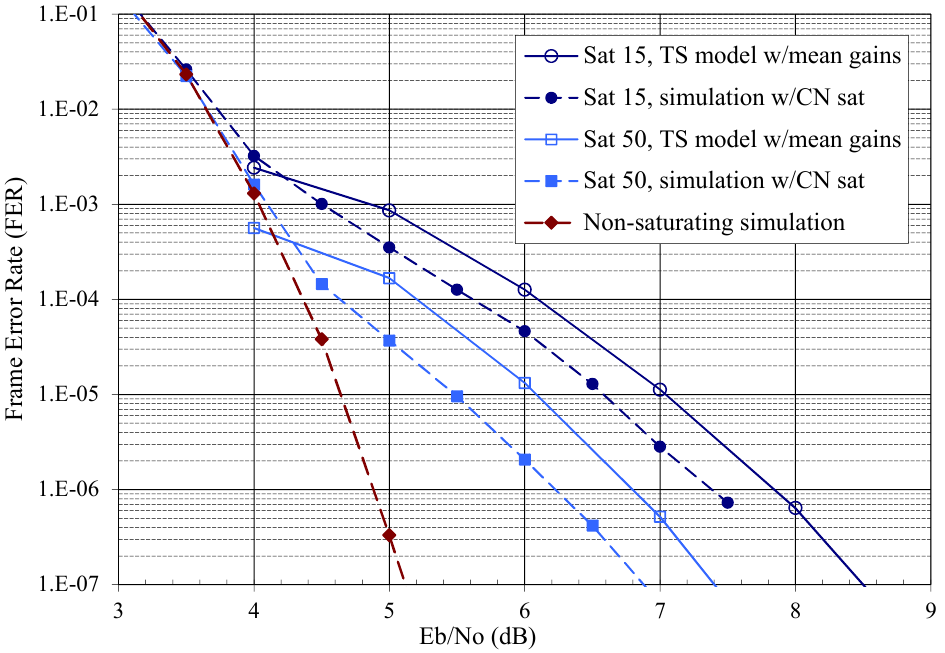}
\caption{FER vs. $\Eb/N_0$ in dB for the $(640,192)$ QC code with simulation and the TS model for predicting error floors using mean gains in the model and no check node inversions.}
\label{fig_Ryan2}
\end{figure}

\begin{figure}
\centering
\includegraphics[width=\figwidth]{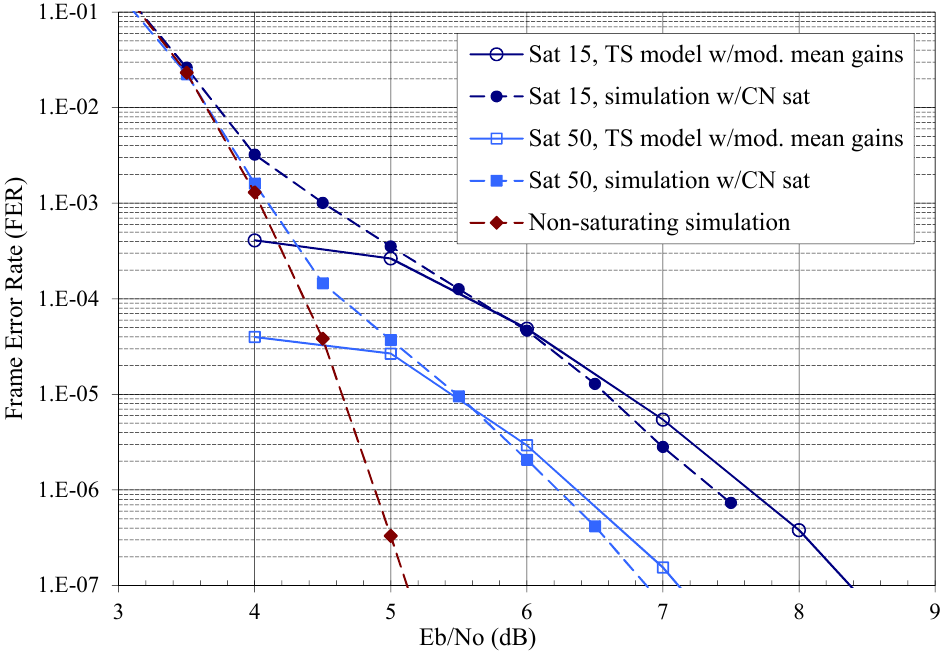}
\caption{FER vs. $\Eb/N_0$ in dB for the $(640,192)$ QC code by simulation and the TS model for predicting error floors using the modified mean gains of \eqref{cinv3} in the model which takes into account check node inversions.}
\label{fig_Ryan3}
\end{figure}

The first code we examine is the $(640,192)$ quasi-cyclic (QC) LDPC code from \cite{Han09}, with $\dv = 5$ and irregular check degree.
The dominant TS for a saturating LLR-domain SPA decoder is the $(5,5)$ elementary TS shown in \cite{Han09}.
The $\matr{A}$ matrix for this TS is primitive and has maximum eigenvalue $r=3$.
The multiplicity of this TS in the code is just $64$ and all such sets have disjoint sets of variable nodes.
Figs.~\ref{fig_Ryan1}, \ref{fig_Ryan2} and \ref{fig_Ryan3} show FER results for simulation and error floor prediction
for this code at two levels of LLR saturation. Our error floor prediction model shows very fast
convergence, attributed to the large eigenvalue (\textit{i.e.}, $r \gg 1$), so we stop the model at $14$ iterations.

The FER simulation results shown in these figures are produced by our LLR-domain SPA decoder,
running on a computer, implemented in double-precision floating-point.
This SPA decoder is non-saturating, in that computations are organized such that saturation will not occur (see \textsection\ref{ss-SPAwo}).
We have not witnessed our non-saturating decoder fail on the $(5,5)$ TSs, which were reported to dominate in \cite{Han09}.
Next, saturation was intentionally introduced into our decoder at levels shown in the figures' legends.
The intentional saturation was introduced at a point corresponding to the output of the complete check node update.
We simulate the transmission of the all-zero codeword for simplicity and take any other decoding result to be decoding failure or frame error.
The simulations were run for a maximum of $200$ iterations.

Fig.~\ref{fig_Ryan1} shows the error floors predicted by (\ref{UnionFER}), but with all check node gains forced to unity.
Fig.~\ref{fig_Ryan2} shows the same conditions, but now with the mean gains of (\ref{gmean2}) applied to the model.
Fig.~\ref{fig_Ryan3} again shows the same conditions, but this time with the modified mean gains of (\ref{cinv3}) applied to 
account for polarity inversions within the degree-two check nodes. 
Comparing the first two figures, one notices that the mean gains provide a significant improvement in the model's accuracy, to within $0.5$ dB of simulation.
Comparing Figs.~\ref{fig_Ryan2} and \ref{fig_Ryan3}, one notices that 
the channel polarity model represents an improvement in the model's accuracy to within $\pm0.1$ dB of simulation. 

Heuristically, the reason that error floor predictions drop when check node gains are introduced has to do with incorrect state buildup in early iterations.
With unity gains, incorrect channel inputs may propagate rapidly in the highly interconnected states of the TS.
With small gains in early iterations this is held off for several iterations, giving the unsatisfied check nodes a better chance to correct the system.
As we previously noted, during early iterations polarity inversions through the degree-two check nodes act to effectively lower the check node gain.
Therefore, as we would expect, introducing the inversions lowers the predicted error floor.

The SPA decoder results presented in \cite{Han09} would appear slightly above the simulation curves with saturation set to $15$ shown in 
Figs.~\ref{fig_Ryan1} to \ref{fig_Ryan3}.
The FER ratio between our non-saturated SPA simulation and the SPA simulation of \cite{Han09} is three orders of magnitude at an
$\Eb/N_0$ of $5$ dB and growing very rapidly.

\begin{figure}
\centering
\includegraphics[width=\figwidth]{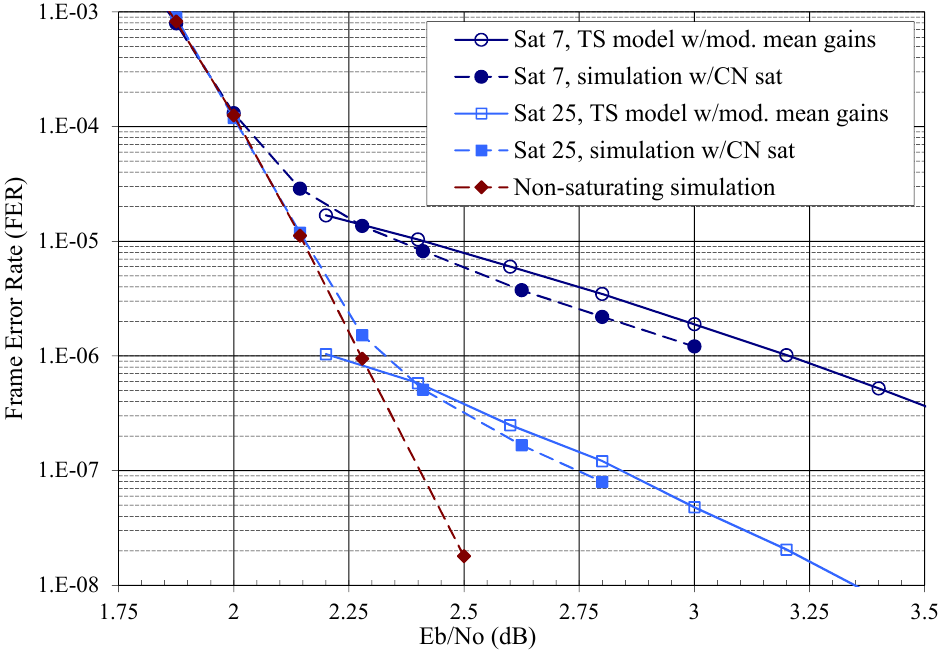}
\caption{FER vs. $\Eb/N_0$ in dB for $(2640, 1320)$ Margulis code with simulation and TS model predicted floor using modified mean gains.}
\label{fig_MargPred}
\end{figure}

The second code analyzed is the $(2640,1320)$ Margulis LDPC code.
Recall that it is a $(3,6)$-regular code with dominant TSs reported to be the $(12,4)$ and $(14,4)$ elementary TSs 
\cite{Han09,RichFloors,MKfloor}.
The $\matr{A}$ matrices for these two TSs are both imprimitive, with $h=2$,
and have maximum eigenvalues $r$ of $1.6956$ and $1.7606$, respectively.
We ran the model for $16$ iterations.
The multiplicity of each of these TSs is $1320$ with lots of overlap between variable nodes of differing sets.
In fact, each $(14,4)$ TS contains all $12$ variable nodes and $2$ of the unsatisfied check nodes of a $(12,4)$ TS, as can
be seen in Fig.~\ref{margtrap}.
Thus, when we compute the failure probability of one $(12,4)$ TS, it largely includes the failure probability of the $(14,4)$ TS that contains it.
Therefore, Fig.~\ref{fig_MargPred} shows the error floors predicted for just the $(12,4)$ TSs, utilizing the model with modified mean gains.
We applied the inversion model at every iteration. 
We find that the state-space model overestimates the simulation's error floor by $0.15$ dB or less.

\begin{figure}
\centering
\includegraphics[width=\figwidth]{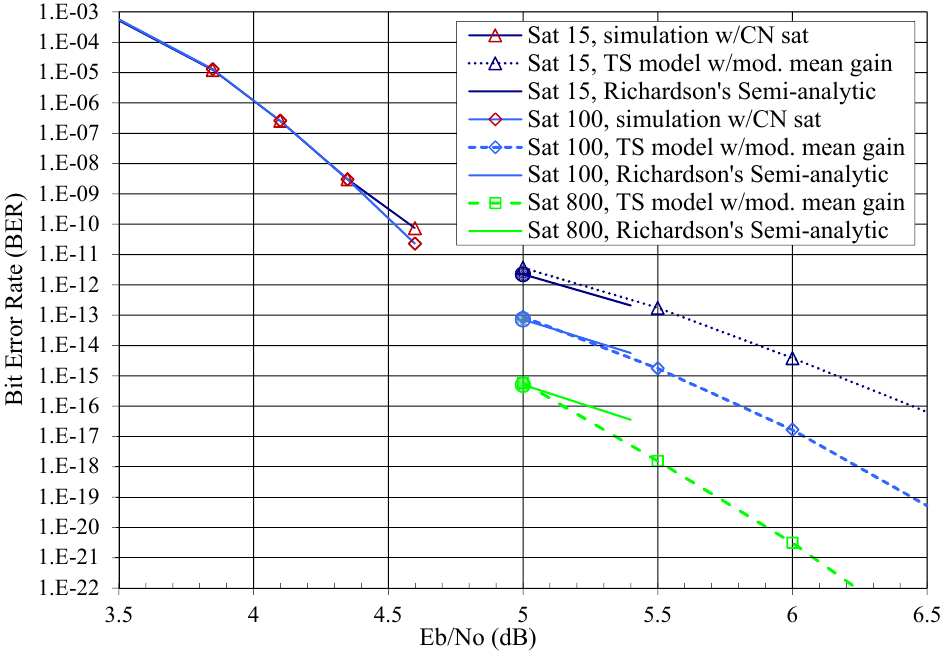}
\caption{BER vs. $\Eb/N_0$ in dB for $(2048, 1723)$ LDPC code from 802.3an with simulation, Richardson's semi-analytical $(8,8)$ floor estimation, and TS model predicted $(8,8)$ floor using modified mean gains.}
\label{fig_802Pred}
\end{figure}

The third code analyzed is the $(2048, 1723)$ Reed-Solomon based LDPC code \cite{DjurRS} from the 802.3an standard.
It is a $(6,32)$-regular code with dominant TSs reported to be the $(8,8)$ elementary TS \cite{SCH10,ZhangQuantFloor}.
The $\matr{A}$ matrix for this TS is primitive, with a maximum eigenvalue $r=4$.
The linear model for this TS converges very fast, in just $3$ iterations when saturation is set to $15$ and a few more iterations at higher saturation limits.
We conservatively ran the model for $16$ iterations, including the inversions at every iteration.
The multiplicity of the $(8,8)$ TS is $14\,272$ \cite{SCH10}.
Fig.~\ref{fig_802Pred} shows the predicted BER error floors of this code utilizing the model with modified mean gains if we assume there are no other TSs
present in this code.

Since a standard Monte Carlo simulation of this low error floor takes a prohibitively large amount of CPU time, we have used Richardson's semi-analytical technique
to estimate the error floor \cite{RichFloors}.
We have run this semi-analytical technique, which was introduced in Section~\ref{sect-margrich}, at an $\Eb/N_0$ of $5.0$ dB (the solid circles on Fig.~\ref{fig_802Pred}) and used Richardson's method of extrapolating the results to $5.4$ dB (the solid lines).
We have only configured the semi-analytical technique to measure the $(8,8)$ TSs.
For the saturation levels of $15$, $100$, and $800$, Richardson's technique required $0.9$, $9.5$, and $1300$ hours of CPU time, respectively.
We used less than $2$ minutes of CPU time to collect the $1000$-frames worth of LLR statistics that we used to produce each error floor prediction point.
Our model's predictions appear to have an accuracy of better than $0.1$ dB at $5.0$ dB.

\begin{figure}
\centering
\includegraphics[width=\figwidth]{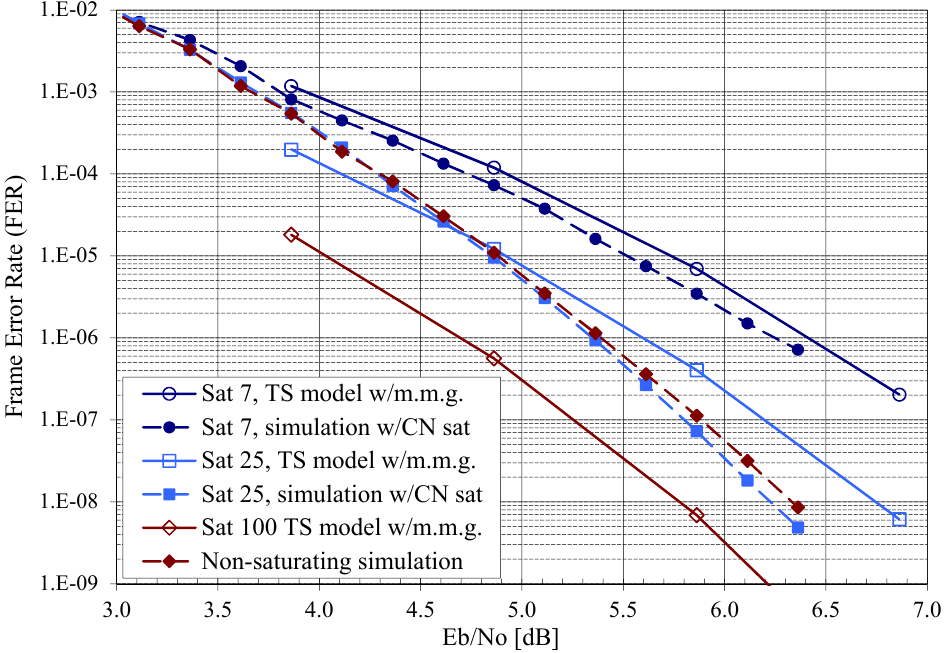}
\caption{FER vs. $\Eb/N_0$ in dB for $(155, 64)$ QC-LDPC code with simulation and TS model predicted floor using modified mean gains.}
\label{fig_155Tan}
\end{figure}

The fourth code analyzed is the $(155, 64)$ QC-LDPC code by Tanner et al. \cite{Tan04}.
It is a $(3,5)$-regular code with the dominant TS reported to be an $(8,2)$ elementary TS, with multiplicity $465$ \cite{SCH11,SCH12}, despite the presence of several other small TSs.  
Additionally, Declercq et al. worked to reduce the floor of this code for the BSC channel in \cite{Declercq155}.
We are interested in understanding if the model breaks down for a code of such short block length.
The $\matr{A}$ matrix for this TS is primitive, with a maximum eigenvalue $r=1.7870$.
Fig.~\ref{fig_155Tan} shows that the predicted FER error floors of this code utilizing the $(8,2)$ model with modified mean gains agrees with simulation to within $0.2$ dB at an LLR saturation level of $7$. 
At this level of saturation an error floor is created; for $\Eb/N_0$ values of at least $4.86$ dB, we measure $69\%$ percent of the failures to be the $(8,2)$ TSs and most of the remaining failures to be other small TSs.
However, at an LLR saturation level of $25$, the error floor appears to be overcome since we measure only $6.8\%$ of the failed frames are on $(8,2)$ TSs.
Given this small fraction of $(8,2)$ failures captured in simulation, we must say that the linear model overestimated TS failures by about $0.9$ dB at $10^{-7}$.

Our findings lead us to remark that the error rate results for this $(155, 64)$ code in \cite{SCH11,SCH12} may be misleading.
In that work an error floor is plotted at a saturation level of $100$ based only on the $(8,2)$ TSs in an SNR region where we believe they will only cause a small fraction of the overall errors observed in practice ($5$ to $7\%$ perhaps).
In fact, we find that some $55\%$ of the failed frames at a saturation level of $100$ contain $11$ or more symbols in error.
Another odd feature of Fig.~\ref{fig_155Tan} is that for the simulation without a saturation limit, the FER actually degrades a little compared to a saturation level of $25$.
Although it is not shown, the simulation of an LLR saturation level of $100$ overlaps the non-saturating simulation.
We attribute the oddities of this example to the shallow nature of waterfall region of the performance curve.

We chose these four LDPC codes as they are quite different and their error floors have appeared in prior studies.
The TSs in these codes have a broad range of multiplicities and degree of overlap among them.
Interestingly, the Margulis code has two dominant TSs, both with imprimitive $\matr{A}$ matrices,
while the other codes are dominated by single TSs with primitive $\matr{A}$ matrices.
Regardless, we find that the state-space model with modified mean gains estimates the simulation's error floor reasonably well and very quickly.


Surprisingly, we find that there are situations in which LLR saturation at the fairly high values of $25$, $50$ and $100$ produce noticeable error floors in simulation and model prediction.
Finally, we note that the terms in (\ref{Qdenom}) containing $\sigma^{2}_{i}$ have a small impact on the overall results.

\section{Conclusion} 

We have drawn into question the error floor levels published in \cite{MKfloor,RichFloors,Han09} and their root cause.
We have shown that the error floor levels are caused by the interaction of non-codeword trapping sets and the numerical effects
of handling highly-certain messages in a belief propagation decoder, such as the LLR-domain SPA decoder.
We have shown that when care is taken in processing those messages the error floor level may be lowered by many orders of magnitude.

We have added some clarity to the situation of two distinctly different applications of the linear system model.
On one hand, Sun introduced the state-space model and proved that many error floors do not exist for infinite-length codes without cycles \cite{SunPhD}.
On the other hand, Schlegel and Zhang improved upon the model and predicted nonzero error floors 
due to the dominant $(8,8)$ trapping sets of the 802.3an code \cite{SCH10}.

With respect to Sun's work, we are in agreement with Sun's conclusion as applied to variable-regular LDPC codes without cycles.
We have shown empirically that the assumptions in the probability of error model do not hold for codes with cycles.
For these codes, we showed that the effects of positively correlated LLRs when allowed to grow (without saturation) drive
the ratio of the mean to standard deviation to a finite value possibly implying an error floor.
This error floor level is difficult to estimate due to the positively correlated LLRs 
and the substantially non-Gaussian distribution of our error indicator.

We are also able to put Schlegel and Zhang's work into perspective.
While not emphasized, their error rate results showing a nonzero error floor for a trapping set were for 
a decoder that saturated (\textit{i.e.}, ``clipped'') the LLR values.
They made no claims with regard to the effects of removing saturation in \cite{SCH10}.
Their addition of mean gains to the model makes a substantial improvement to error floor prediction as we have shown.
Our realization of the model is more general than theirs, in that we do not require that $\matr{B}=\matr{B}_{\mathrm{ex}}$, 
that $\matr{B}$ has regular column weight, and that $\matr{A}$ has regular row and column weight\footnote{If $\matr{A}$ has regular row or column weight, then its spectral radius $\rho(\matr{A})$ is integral.}.
Additionally, we have simplified considerably the incorporation of polarity inversion by the degree-two check nodes into the model.
We have presented two new methods to collect the statistics needed to drive the model.
We have shown that the reason that the model predicts saturated performance relatively well is that by the iteration count at which LLRs
become substantially correlated, their variance is driven to zero by the effect of saturation.
This means that the model derived for the case with no cycles actually works rather well for the case with cycles when there is a significant saturation effect present.
In our experiments the model with modified mean gains estimated the FER floor of floating-point simulation to within $0.2$ dB for several different LDPC codes.

Altogether, we have reached a better understanding of the trapping set failure mechanisms.
Trapping set failures can be frequently corrected when the beneficial unsatisfied check LLRs eventually overpower the trapping set's own detrimental LLR growth.
Achieving this reliably can take significant LLR metric dynamic range.
The greater the LLRs are allowed to grow the more often channel inversions aligned to trapping sets can be overcome.
Chen \textit{et al.} had the insightful observation in \cite{ChenFoss},
``...that the error floor is caused by the combined effects of short cycles in the graph representation of the code and of clipping.''

This phenomenon also helps explain prior observations.
Others have noted that ``stable'' trapping set failures converge very rapidly (\textit{cf.} \S15.3.1 of \cite{RyanLin}), often in less than ten iterations.
We now realize that this is due to the saturation point of the beneficial unsatisfied check nodes being reached and bringing the correction process to a halt.
When it is reached and the states are still in error the structure is very stable when it is an absorbing set.
Absorbing sets guarantee that unsatisfied check messages are always outnumbered by internal messages at any variable node within the set.
Metric saturation is why the bit-flipping decoder analogy was made in \cite{DolecekTIT} to explain absorbing sets.
The process of passing real-valued messages in an LLR-domain SPA decoder degenerates to passing positively or negatively saturated messages
where the outcomes can be explained by bit-flipping decoder logic.
We have shown that when saturation degenerates the SPA decoder to a bit flipping decoder, it actually helps to correct variable nodes in 
the trapping set until an absorbing set is reached.
We have used this observation to modify Richardson's semi-analytical technique so that it may estimate error floors of a non-saturating SPA decoder.

The results make clear that when presenting error floors, future researchers should document the levels of any LLR saturation.

Future work may include extensions to more channels, to non-elementary trapping sets, and to account for correlations among LLRs.
We hope that many of the techniques covered in this paper generalize to variable-irregular codes, but we recognize 
that the issues of bounding and approximating the trapping sets' spectral radii must be reconsidered.

\appendices
\section{Reducible $\matr{A}$ Matrices} 
\label{app-red}
In this Appendix we address the special case of reducible $\matr{A}$ matrices.

\begin{lem}
\label{Lcyclic}
Let $G$ be a connected multigraph of order $n \ge 2$.
Then $G$ is isomorphic to the cycle graph $C_n$ if and only if every vertex is degree two.
\end{lem}
\begin{IEEEproof}
Omitted.
\end{IEEEproof}

\begin{thm}
Consider a connected multigraph $G$ which meets Assumption~\ref{assg2}.
$G$ is either a cycle graph or has an associated $\matr{A}(D)$ matrix that is irreducible.
\end{thm}
\begin{IEEEproof}
Multigraphs with any vertices of degree one are not allowed by Assumption~\ref{assg2}.
Connected multigraphs with all vertices of degree two are cyclic by Lemma~\ref{Lcyclic}.
So now we will show any possible remaining connected multigraphs must have an irreducible adjacency matrix $\matr{A}(D)$.

The remaining multigraphs must have at least one vertex of degree greater than two.
Also, all vertex degrees must sum to an even number, as every edge joins two vertices.
This yields a connected multigraph with at least two cycles. Since a walk without backtracking through a connected multigraph
with two cycles is able to reverse direction, every edge may be visited in either direction.
Thus, when the edges of $G$ are expanded to be vertices of the digraph $D$, $D$ will be strongly connected.
Since $D$ it is strongly connected, its associated adjacency matrix $\matr{A}(D)$ will be irreducible by Lemma~\ref{Lconn}.
\end{IEEEproof}

Of course, since $\matr{A}(D)$ is irreducible so must be $\matr{A}(D)^T$, which we use as $\matr{A}$ in our state-space model.

Therefore, the remainder of this appendix need only address the properties of cycle graphs.
The expansion of the edges of cycle graph $G=C_a$ to a digraph $D$ will produce two disconnected directed cycles
of length $a$, one associated with the clockwise non-backtracking walk of $G$ and one with the counter-clockwise non-backtracking walk of $G$.
Thus, $\matr{A}(D)$ may be symmetrically permuted to 
$\matr{P} \matr{A}(D) \matr{P}^T  = \bigl[\begin{smallmatrix} \matr{A}_1&\matr{0}\\\matr{0}&\matr{A}_2\end{smallmatrix}\bigr]$, 
in which both $\matr{A}_1$ and $\matr{A}_2$ are irreducible.
Since the component digraphs are each directed cycles of length $a$, for some $\matr{P}$, 
both adjacency submatrices $\matr{A}_1$ and $\matr{A}_2$ are $a\times a$ permutation matrices that implement a circular shift by one position.
Thus, both $\matr{A}_1$ and $\matr{A}_2$ are imprimitive with $h=a$.
Since every row and column of $\matr{A}_1$ and $\matr{A}_2$ have weight one, 
all their eigenvalues lie on the unit circle, $r=\rho(\matr{A}_1)=$ $\rho(\matr{A}_2) = 1$.
The eigenvectors of $\matr{A}_1$ and $\matr{A}_2$ must be identical.
The eigenvectors of $\matr{A}_1$ may be enlarged to be the eigenvectors of $\matr{A}$ by appending $a$ zeros.
As $\matr{A}_1$ is a permutation matrix, the eigenvalue $r=1$ is associated
with the all-one eigenvector (or a scalar multiple thereof): $\matr{A}_1 \vect{v}_{1,1} = r \: \vect{v}_{1,1}$,
where $\vect{v}_{1,1} = [1,1, \ldots 1]^T$.
By a similar argument the associated left eigenvector of $\matr{A}_1$ is $\vect{w}_{1,1} = [1,1, \ldots 1]^T$.
Similar statements hold for $\matr{A}_2$.

Define $\vect{\tilde w}_{1,1}$ to be $\vect{w}_{1,1}$ appended with $a$ zeros, and $\vect{\tilde w}_{2,1}$
to be $\vect{w}_{2,1}$ prefixed with $a$ zeros.
Now, let us complete the derivations in Section~\ref{sect-domprob} using $\beta_l \triangleq \vect{ w}_1^T \vect{x}_l $,
with $\vect{ w}_1 \triangleq \vect{\tilde w}_{1,1} + \vect{\tilde w}_{2,1}$.
Since $\vect{\tilde w}_{1,1}$ and $\vect{\tilde w}_{2,1}$ are associated with the same eigenvalue,
their sum $\vect{w}_1 = [1,1, \ldots 1]^T$ is also a left eigenvector of $\matr{A}$.
Note that $\vect{w}_1$ is not the only positive left eigenvector of $\matr{A}$, as any linear combination
$c_1 \vect{\tilde w}_{1,1} + c_2 \vect{\tilde w}_{2,1}$ with $c_i>0$ would also be a positive left eigenvector.
For the cycle graphs under consideration, (\ref{beta2}) becomes
\begin{equation}
\begin{split}
\beta_l^{'} &= \left( \vect{\tilde w}_{1,1}^T + \vect{\tilde w}_{2,1}^T \right) \matr{B} \vlambda \prod_{j=1}^l g_j \\
&+ \sum_{i=1}^{l} \left( \vect{\tilde w}_{1,1}^T + \vect{\tilde w}_{2,1}^T \right) \left( \matr{B} \vlambda + \matr{B}_{\mathrm{ex}} \vlambda_i^{(\mathrm{ex})} \right) \prod_{j=i+1}^l g_j.
\end{split}
\end{equation}

This is equivalent to the other expressions derived in Section~\ref{sect-domprob} since $\vect{ w}_1 = \vect{\tilde w}_{1,1} + \vect{\tilde w}_{2,1}$, noting that $r=1$ here.
With these considerations we may use the results of Section~\ref{sect-domprob} and later for the allowed reducible $\matr{A}$ matrices.
Relative to the variable-regular codes addressed in this \paperchapt/, the cycle graph $C_a$ corresponds to the $(a,b)$ trapping set with $a\ge2$ and $b=(\dv-2)a$.

\section{Adding Leaves and Branches} 
\label{app-add}

We now expand the set of multigraphs addressed in this work to include those with leaves and branches, which were
previously eliminated by Assumption~\ref{assg2}.
Thus, the variable nodes in the associated Tanner subgraphs will be permitted to have $\dv-1$ adjacent degree-one check nodes.
In \cite{SunPhD}, Sun referred to the graphs addressed in this appendix as ``trapping sets with external data nodes.''

We will describe the elementary Tanner graphs of this appendix from the perspective of their
associated multigraphs using the mapping of Lemma~\ref{gmap}.
Let a \emph{base} graph be a multigraph that meets Assumptions~\ref{ass_elem}, \ref{assg1}, and \ref{assg2}.
Such a graph contains one or more cycles.
A \emph{leaf} is a vertex of degree one.
A \emph{branch} is a vertex of degree two or more outside of the base graph, and is not contained in any cycles.

\begin{exmp}
Fig.~\ref{fig_inv1} contains one leaf on a base graph and Fig.~\ref{fig_inv2} contains just three leaves and no base graph.
The $(18,8)$ failing structure of Fig.~\ref{188trap} has leaves A, B, and D and branch C on a base graph that is the 
$(14,4)$ trapping set of Fig.~\ref{144trap}.
\end{exmp}

\begin{ass}[Replacement for Assumption \ref{assg2}]
\label{assgn}
Multigraphs of interest must contain a base graph and zero or more leaves and branches.
\end{ass}

We require a base graph because without a base graph the eigenvalues would all be zero, as will become apparent shortly.
Our new assumption allows the variable nodes within the associated Tanner subgraphs to neighbor up to $\dv-1$ degree-one check nodes.

\begin{thm}
\label{zeig}
Let $G=(V,E)$ be a multigraph meeting Assumptions~\ref{ass_elem}, \ref{assg1}, and \ref{assgn} containing base graph $G_B$, $n$ leaves, with $n\ge 1$,
and zero or more branches.
Let $D=(Z,A)$ and $D_B$ be the associated digraphs, created as described in Section~\ref{ss-relat}.
Then, the adjacency matrix $\matr{A}(D)$ is reducible and its spectrum contains the eigenvalues of $\matr{A}(D_B)$ and zeros.
\end{thm}
\begin{IEEEproof}
Each additional leaf in $G$ adds one edge to $G$. 
Let the edge $e_i\in E$ join leaf $v_k$ to vertex $v_j$ in G, where $v_j$ is not a leaf.
The edge $e_i$ maps to vertices $z_i, z_i{'} \in Z$ in $D$.
These vertices are not strongly connected to the digraph as $d^+_i=0$ and $d^-_{i'}=0$, and hence $\matr{A}(D)$ must be reducible by Lemma~\ref{Lconn}.

Every leaf in $G$ creates an all-zero column in $\matr{A}(D)$ due to $d^-_{i'}=0$ and an all-zero row due to $d^+_i=0$.
The all-zero columns may be symmetrically permuted to the left (\textit{i.e.}, their corresponding vertices are relabeled with the lowest possible values)
and the all-zero rows to the bottom 
to form
\begin{equation}
\label{Aleaves}
\matr{P}\, \matr{A}(D)\, \matr{P}^T = \begin{bmatrix}
\matr{0}&\matr{Y}_1&\matr{Y}_2\\
\matr{0}&\matr{B}&\matr{Y}_3\\
\matr{0}&\matr{0}&\matr{0}%
\end{bmatrix},
\end{equation}
where the block diagonal contains square submatrices.
The $n\times n$ zero matrices at both ends of the block diagonal correspond to the $n$ leaves in $D$.

The eigenvalues of $\matr{A}(D)$ are the roots of the characteristic equation $\det(\matr{A}(D) -\mu \matr{I})=0$,
which simplifies to $\det(\matr{B} -\mu \matr{I})\mu^{2n}=0$ by use of the expansion by minors along every zero row and column.
Thus, the eigenvalues of $\matr{A}(D)$ are the eigenvalues of $\matr{B}$ and $2n$ zeros.
In case $G$ contains both leaves and branches, the removal of one layer of leaves exposes a new layer of leaves and the
operations in this paragraph must be repeated until we work down to $\matr{B}=\matr{A}(D_B)$, which will be irreducible
except for the case addressed in Appendix~\ref{app-red} in which $G_B$ is a cycle graph.
\end{IEEEproof}

By showing that the nonzero eigenvalues are preserved by the addition of leaves and branches, it is
simple to argue that Theorem~\ref{rdv}, which bound the spectral radius of $\matr{A}(D)$,  
still hold if Assumption~\ref{assg2} is replaced by \ref{assgn}.

The only weakness with respect to our prior development is that we can no longer assume that the left eigenvector $\vect{w}_1$ is positive as $\matr{A}(D)$ 
may now be reducible. 
In practice, we have found that the new entries added to $\vect{w}_1$ by the addition of leaves and branches are half zero and half positive.
This may be proved by applying the Subinvariance Theorem in \cite[p.~23]{Seneta} to (\ref{Aleaves}).
The presence of zero entries in $\vect{w}_1$ weakens our prior claims on the error indicator $\beta_l \triangleq \vect{w}_1^T \vect{x}_l$.
Now, the error indicator is most effective on the variable nodes corresponding to the base graph and less effective on
the variable nodes corresponding to the branches and leaves.

Finally, we do not see much practical motivation to predict the error floors of graphs with branches and leaves.
We would prefer to run predictions on the base graphs contained within.
Since the graphs with branches and leaves have the same spectral radius as their base graph they are no more
likely to fail in theory.  
In fact, they should be less likely to fail as more channel values are involved and more unsatisfied check nodes are working
to correct the graph.

\section{Tables of Absorbing Set Structure} 
\label{app-tab}
As opposed to studying the trapping sets that dominate several specific codes,
in this appendix we present an overview of all subgraphs that may become troublesome trapping
sets which meet specific conditions for particular large classes of codes.
This can serve to put some perspective on the structural parameters that
have been discussed in this \paperchapt/ such as: $a$, $b$, the index of imprimitivity $h$, and the spectral radius $r$.

The variable-regular codes we examine are assumed to be described by Tanner graphs that are $4$-cycle-free.
We divide the information into tables by their variable-degree $\dv$.
The further conditions we put on the subgraphs of interest are described in Assumptions~\ref{ass_elem} and \ref{assg1} and Definition \ref{def_absorbing};
the subgraphs must be elementary, connected and absorbing, respectively.

\begin{table}[!t]
\caption{Connected, elementary absorbing sets from the set of $4$-cycle-free $\dv = 3$ variable-regular codes, with $r_{\max} > 1.3$}
\label{table_3dv}
\centering
\begin{tabular}{c||r||r|r|r}
\hline
$(a,b)$ & Num. & $h_{\max}$ & $r_{\min}$ & $r_{\max}$ \\
\hline\hline
4,0  &	1 &	1 &	2       &	2 \\
4,2  &	1 &	1 &	1.521 &	1.521 \\
5,1  &	1 &	1 &	1.829 &	1.829 \\
5,3  &	2 &	4 &	1.414 &	1.424 \\
6,0  &	2 &	2 &	2       &	2 \\
6,2  &	4 &	2 &	1.696 &	1.729 \\
6,4  &	4 &	2 &	1.348 &	1.361 \\
7,1  &	4 &	1 &	1.883 &	1.888 \\
7,3  &	10 &	2 &	1.599 &	1.665 \\
7,5  &	6 &	2 &	1.298 &	1.316 \\
8,0  &	5 &	2 &	2       &	2 \\
8,2  &	19 &	2 &	1.780 &	1.870 \\
8,4  &	25 &	2 &	1.521 &	1.622 \\
9,1  &	19 &	1 &	1.911 &	1.929 \\
9,3  &	63 &	2 &	1.696 &	1.851 \\
9,5  &	52 &	2 &	1.463 &	1.592 \\
10,0 &19  &	2 &	2 	&	2 \\
10,2 &113 &	2 &	1.829 &	1.911 \\
10,4 &198 &	2 &	1.629 &	1.841 \\
10,6 &109 &	4 &	1.414 &	1.570 \\
11,1 &114 &	1 &	1.929 &	1.947 \\
11,3 &482 &	2 &	1.758 &	1.899 \\
11,5 &536 &	2 &	1.571 &	1.836 \\
11,7 &197 &	2 &	1.379 &	1.555 \\
12,0  &	85 &	2 &	2 &	2 \\
12,2  &	835 &	2 &	1.861 &	1.940 \\
12,4  &	1,892 &	2 &	1.696 &	1.894 \\
12,6  &	1,373 &	2 &	1.521 &	1.833 \\
12,8  &	351 &	2 &	1.348 &	1.545 \\
13,1  &	839 &	1 &	1.941 &	1.961 \\
13,3  &	4,541 &	2 &	1.799 &	1.934 \\
13,5  &	6,374 &	2 &	1.645 &	1.891 \\
13,7  &	3,159 &	2 &	1.481 &	1.831 \\
13,9  &	581 &	2 &	1.321 &	1.537 \\
14,0  &	509 &	2 &	2 &	2 \\
14,2  &	7,589 &	2 &	1.883 &	1.954 \\
14,4  &	21,434 &	2 &	1.745 &	1.931 \\
14,6  &	19,587 &	2 &	1.599 &	1.889 \\
14,8  &	6,879 &	2 &	1.446 &	1.830 \\
14,10  &	931 &	4 &	1.298 &	1.532 \\
\hline
\end{tabular}
\end{table}

\begin{table}[t]
\caption{Connected, elementary absorbing sets from the set of $4$-cycle-free $\dv = 4$ variable-regular codes}
\label{table_4dv}
\centering
\begin{tabular}{c||r||r|r|r}
\hline
$(a,b)$ & Num. & $h_{\max}$ & $r_{\min}$ & $r_{\max}$ \\
\hline\hline
4,4  &	1 &	1 &	2 &	2 \\
5,0  &	1 &	1 &	3 &	3 \\
5,2  &	1 &	1 &	2.629 &	2.629 \\
5,4  &	1 &	1 &	2.219 &	2.219 \\
6,0  &	1 &	1 &	3 &	3 \\
6,2  &	2 &	1 &	2.697 &	2.710 \\
6,4  &	3 &	1 &	2.355 &	2.367 \\
6,6  &	2 &	2 &	2 &	2 \\
7,0  &	2 &	1 &	3 &	3 \\
7,2  &	7 &	1 &	2.744 &	2.762 \\
7,4  &	11 &	2 &	2.449 &	2.480 \\
7,6  &	4 &	1 &	2.159 &	2.160 \\
8,0  &	6 &	2 &	3 &	3 \\
8,2  &	28 &	2 &	2.778 &	2.805 \\
8,4  &	50 &	2 &	2.525 &	2.585 \\
8,6  &	28 &	2 &	2.272 &	2.296 \\
8,8  &	5 &	2 &	2 &	2 \\
9,0  &	16 &	1 &	3 &	3 \\
9,2  &	126 &	1 &	2.805 &	2.850 \\
9,4  &	285 &	2 &	2.584 &	2.728 \\
9,6  &	177 &	2 &	2.355 &	2.429 \\
9,8  &	27 &	1 &	2.126 &	2.141 \\
10,0  &	59 &	2 &	3 &	3 \\
10,2  &	719 &	2 &	2.826 &	2.886 \\
10,4  &	1,915 &	2 &	2.629 &	2.821 \\
10,6  &	1,404 &	2 &	2.422 &	2.648 \\
10,8  &	298 &	2 &	2.219 &	2.313 \\
10,10  &	19 &	2 &	2 &	2 \\
11,0  &	265 &	1 &	3 &	3 \\
11,2  &	4,721 &	1 &	2.843 &	2.903 \\
11,4  &	14,569 &	2 &	2.667 &	2.852 \\
11,6  &	12,458 &	2 &	2.478 &	2.788 \\
11,8  &	3,231 &	2 &	2.295 &	2.458 \\
11,10  &	208 &	1 &	2.104 &	2.127 \\
\hline
\end{tabular}
\end{table}

\begin{table}[t]
\caption{Connected, elementary absorbing sets from the set of $4$-cycle-free $\dv = 5$ variable-regular codes, with $r_{\max} > 2.6$}
\label{table_5dv}
\centering
\begin{tabular}{c||r||r|r|r}
\hline
$(a,b)$ & Num. & $h_{\max}$ & $r_{\min}$ & $r_{\max}$ \\
\hline\hline
5,5  &	1 &	1 &	3 &	3 \\
5,7  &	1 &	1 &	2.629 &	2.629 \\
6,0  &	1 &	1 &	4 &	4 \\
6,2  &	1 &	1 &	3.693 &	3.693 \\
6,4  &	2 &	1 &	3.360 &	3.403 \\
6,6  &	4 &	1 &	3 &	3.112 \\
6,8  &	5 &	1 &	2.697 &	2.767 \\
7,1  &	1 &	1 &	3.875 &	3.875 \\
7,3  &	5 &	1 &	3.596 &	3.669 \\
7,5  &	14 &	1 &	3.307 &	3.427 \\
7,7  &	23 &	1 &	3 &	3.130 \\
7,9  &	25 &	1 &	2.744 &	2.827 \\
8,0  &	3 &	1 &	4 &	4 \\
8,2  &	16 &	1 &	3.775 &	3.827 \\
8,4  &	68 &	1 &	3.522 &	3.673 \\
8,6  &	165 &	1 &	3.271 &	3.465 \\
8,8  &	252 &	2 &	3 &	3.208 \\
8,10  &	232 &	2 &	2.778 &	2.918 \\
8,12  &	124 &	2 &	2.525 &	2.619 \\
9,1  &	28 &	1 &	3.904 &	3.905 \\
9,3  &	276 &	1 &	3.693 &	3.769 \\
9,5  &	1,151 &	2 &	3.464 &	3.665 \\
9,7  &	2,541 &	2 &	3.243 &	3.521 \\
9,9  &	3,284 &	2 &	3 &	3.289 \\
9,11  &	2,541 &	2 &	2.805 &	3.071 \\
9,13  &	1,150 &	2 &	2.584 &	2.751 \\
10,0  &	60 &	2 &	4 &	4 \\
10,2  &	1,188 &	2 &	3.822 &	3.870 \\
10,4  &	8,435 &	2 &	3.626 &	3.789 \\
10,6  &	27,706 &	2 &	3.421 &	3.810 \\
10,8  &	49,991 &	2 &	3.220 &	3.717 \\
10,10  &	53,884 &	2 &	3 &	3.440 \\
10,12  &	35,721 &	2 &	2.826 &	3.191 \\
10,14  &	14,308 &	2 &	2.629 &	3.009 \\
10,16  &	3,224 &	2 &	2.422 &	2.652 \\
\hline
\end{tabular}
\end{table}

Rather than find every single graph that meets our parameters we can simplify the search considerably.
As we are just interested in a graph's structure as opposed to its labeling we need only identify
one of the graphs among several that are isomorphic to the others.
This is known as partitioning the graphs into \emph{equivalence classes}.

Graph theory tools such as ``geng'' \cite{nauty} can generate non-isomorphic simple graph descriptions very quickly satisfying sets of parameters such as ours.
We run this tool once for each row of the tables. 
Each time we configure it to find undirected graphs of order $a$, size $(a \dv - b)/2$, and with vertex degrees in 
the interval $\left[ \left\lceil \dv/2 \right\rceil, \dv \right]$.
The size we have specified is a direct result of Euler's handshaking lemma.
By limiting the tool output to simple graphs, we eliminate multigraphs of girth $2$ and their equivalent Tanner subgraphs of girth $4$.
The range of vertex degrees specified ensures that we get absorbing sets.
We then process each graph in a custom tool that converts it to the adjacency matrix of the associated directed graph to find its spectral radius $r$ and index of imprimitivity $h$.

Table~\ref{table_3dv} shows the graphs found based on the set of $4$-cycle-free codes with $\dv = 3$.
The graphical equivalence classes found are divided into rows by their $(a,b)$ parameters listed in the first column.
The second column of each row presents the number of equivalence classes found that meet the $(a,b)$ parameters and all our other assumptions.
The third column shows the maximum index of imprimitivity $h$ over the $(a,b)$ equivalence classes.
The fourth and fifth columns show the minimum and maximum spectral radius $r$ over the $(a,b)$ equivalence classes.
To save space we have left out the weakest $(a,b)$ pairs with $r_{\max} \le 1.3$. 
Tables~\ref{table_4dv} and \ref{table_5dv} present similar results based on codes with $\dv = 4$ and $5$, respectively,
omitting the $(a,b)$ pairs with $r_{\max} \le 2.6$ for $\dv=5$. 

\begin{table}[t]
\renewcommand{\arraystretch}{1.2} 
\caption{Extreme Relative Estimation Errors for the Spectral Radius}
\label{table_specsum}
\centering
\begin{tabular}{c||r|r|r|r}
\hline
\bfseries Originating  & \multicolumn{2}{c|}{\bfseries With respect} & \multicolumn{2}{c}{\bfseries With respect}\\
\bfseries Table  & \multicolumn{2}{c|}{\bfseries to \eqref{specSchlegel} and} & \multicolumn{2}{c}{\bfseries to \eqref{specButler}  and}\\
\cline{2-5}
\bfseries         &\multicolumn{1}{c|}{$r_{\min}$}&\multicolumn{1}{c|}{$r_{\max}$}&\multicolumn{1}{c|}{$r_{\min}$}&\multicolumn{1}{c}{$r_{\max}$}\\
\hline\hline
\ref{table_3dv} ($\dv=3$)    &$0.0\%$ &$-21.9\%$ &$6.1\%$ &$-16.4\%$\\
\ref{table_4dv} ($\dv=4$)    &$0.0\%$ &$-12.0\%$ &$2.1\%$ &$-9.4\%$\\
\ref{table_5dv} ($\dv=5$)    &$0.0\%$ &$-13.9\%$ &$1.0\%$ &$-20.2\%$\rlap{\textsuperscript{a}}\\
\ref{table_5dv} with $b\le a$&$0.0\%$ &$-13.9\%$ &$1.0\%$ &$-12.9\%$\\
\hline
\multicolumn{5}{l}{\footnotesize \textsuperscript{a}Discovered over the relatively weak $(10,16)$ trapping sets.}%
\end{tabular}
\end{table}

These tables present enough information to comment on the approximations to the spectral radius $r$ developed in Section~\ref{ss-relat}.
The extreme relative estimation errors are summarized in Table~\ref{table_specsum}.
We find empirically that the approximation of \eqref{specSchlegel} does not overestimate the spectral radius, but generally underestimates it.
For $90\%$ of the $\dv = 3$ equivalence classes, this approximation deviates by no more than $6.0\%$ below the true value of the spectral radius.
The approximation of \eqref{specButler} generally produces larger estimates of the spectral radius, sometimes overestimating the true value.
For the $\dv = 3$ equivalence classes shown in Table~\ref{table_3dv}, the mean approximation error is $-3.5\%$ for \eqref{specSchlegel} and $+1.3\%$ for \eqref{specButler}.  
For both estimators, we measure a standard deviation of the relative error of about $2.0\%$.  
Since we find that the correlation coefficient between the relative errors of these two estimates to be $0.732$, we may reduce the standard deviation and bias by combining them.
In summary, it appears that \eqref{specButler} and the combined estimator are marginally better estimators than \eqref{specSchlegel}.

As discussed in Section~\ref{ss-relat}, the approximation \eqref{specButler} assumes that the trapping set contains mostly zero or one unsatisfied check node per variable node, which is not always true in Table~\ref{table_5dv} (\textit{i.e.}, $\dv=5$).  
When the assumption does not hold, the approximation of \eqref{specButler} can underestimate the spectral radius significantly, as demonstrated by the extreme value within  Table~\ref{table_specsum}.  
In this case, the $-20.2\%$ error occurred over the relatively weak $(10,16)$ trapping sets.
If we limit the processing of Table~\ref{table_5dv} to its rows such that $b\le a$, \textit{i.e.}, the strong trapping sets, then the extreme relative error for this estimator would be reduced to $-12.9\%$ as shown in the final row of Table~\ref{table_specsum}.

\section*{Acknowledgment} 

The authors would like to thank Aravind Iyengar, Christian Schlegel, Roxana Smarandache, and Xiaojie Zhang for their helpful discussions and encouragement.
Further, the authors are indebted to Xiaojie Zhang for verifying several of our low floor results in the AWGN channel 
with an independent implementation of the non-saturating SPA decoder.
Also, we thank Yang Han and William Ryan for providing the parity-check matrix of the $(640,192)$ QC code used in Section~\ref{sect-numer}.



\bibliographystyle{IEEEtran}
\bibliography{IEEEabrv,Butler}


\end{document}